\documentclass[aps,pra,superscriptaddress,amsmath,amssymb,10pt]{revtex4-1}

\usepackage{comment}
\usepackage{tikz}
\usetikzlibrary{trees}
\usetikzlibrary{decorations.pathmorphing}
\usetikzlibrary{decorations.markings}
\tikzset{
    photon/.style={decorate, decoration={snake,segment length=1.5mm}, draw=black},
    coulomb/.style={dotted},
    electron/.style={draw=black, postaction={decorate},
        decoration={markings,mark=at position .55 with {\arrow[draw=black]{>}}}}, 
    gluon/.style={decorate, draw=magenta,
        decoration={coil,amplitude=4pt, segment length=5pt}},
    boundelectron/.style={thick, double},
    transverse/.style={dashed}
}

\usepackage{graphicx}
\usepackage{dcolumn}
\usepackage{bm}
\usepackage{rotating}
\usepackage{dcolumn}

\newcolumntype{.}{D{.}{.}{8}}

\usepackage[utf8]{inputenc}
\usepackage{physics}
\usepackage{amsmath, amssymb}
\usepackage{tabularx,booktabs,array,dcolumn}
\usepackage{setspace}
\usepackage{vmargin}
\usepackage{xcolor}
\usepackage{graphicx,psfrag,subfigure}

\usepackage{float}
\usepackage[all]{xy}

\usepackage{bm}
\usepackage{mathtools}
\usepackage[english]{babel}
\usepackage{siunitx,letltxmacro}
\LetLtxMacro{\svqty}{\qty}
\usepackage{physics}
\LetLtxMacro{\qty}{\svqty}

\newcommand{\bos}[1]{\boldsymbol{#1}}
\newcommand{\mr}[1]{\mathrm{#1}}

\def\Eh{E_\mathrm{h}}

\def\iim{\mr{i}}
\def\eem{\mr{e}}

\def\nb{N_\text{b}} 
\def\naux{N_\text{aux}} 

\def\four{^{[4]}} 

\def\bp{\bos{p}}
\def\br{\bos{r}}

\def\balpha{\bos{\alpha}}
\def\bsigma{\bos{\sigma}}

\def\lphi{\tilde{\phi}}

\def\epsi{\varepsilon}

\def\tC{\text{C}}
\def\tB{\text{B}}
\def\ti{\text{i}}

\def\som{Supplementary Material}

\usepackage{soul}
\def\wp{\text{wp}}
\def\np{\text{np}}
\def\pupu{{++}}
\def\mimi{{--}}
\def\brra{\text{BR}}
\def\mcL{\mathcal{L}}
\def\tildi{{\tilde{i}}}

\newcommand{\brapsket}[4]{\left\langle\begin{array}{@{}c@{}} #1 \\[0.05cm] #2 \end{array} \Big| \begin{array}{c} #3 \\[0.05cm] #4 \end{array} \right\rangle}
\newcommand{\braharom}[3]{\left\langle\begin{array}{@{}c@{}} #1 \\[0.05cm] #2 \\[0.05cm] #3 \end{array} \right. \Bigg| }
\newcommand{\ketharom}[3]{\left. \Bigg | \begin{array}{@{}c@{}} #1 \\[0.05cm] #2 \\[0.05cm] #3 \end{array} \right\rangle }

\def\honehtwo{$h_1h_2$}

\definecolor{ao}{rgb}{0.0, 0.5, 0.0}

\newcolumntype{d}[1]{D{.}{.}{#1}}

\usepackage[unicode]{hyperref}
\hypersetup{
   unicode=true,          
   plainpages=false,
   colorlinks=true,       
   linkcolor=black,          
   linkcolor=blue,          
   citecolor=blue,        
   urlcolor=blue           
}

\usepackage{url}
\begin{document}

\title{%
Double-pair Coulomb and Breit photon correction to the correlated relativistic energy
}

\author{P\'eter Jeszenszki} 
\author{Edit M\'atyus} 
\email{edit.matyus@ttk.elte.hu}
\affiliation{MTA–ELTE Lendület `Momentum' Molecular Quantum electro-Dynamics Research Group,
Institute of Chemistry, Eötvös Loránd University, Pázmány Péter sétány 1/A, Budapest, H-1117, Hungary}

\date{\today}

\begin{abstract}
\noindent %
The simplest, algebraic quantum-electrodynamical corrections due to the double-negative energy subspace and instantaneous interactions are computed to the no-pair energy of two-spin-1/2-fermion systems. 
Numerical results are reported for two-electron atoms with a clamped nucleus and positronium-like genuine two-particle systems. 
The Bethe-Salpeter equation provides the theoretical framework, and numerical methods have been developed for its equal-time time-slice. 
In practice, it requires solving a sixteen-component eigenvalue equation with a two-particle Dirac Hamiltonian, including the appropriate interaction. 
The double-pair corrections can either be included in the interaction part of the eigenvalue equation or treated as a perturbation to the no-pair Hamiltonian. 
The numerical results have an $\alpha$ fine-structure constant dependence that is in excellent agreement with the known  $\alpha^3\Eh$-order double-pair correction of non-relativistic quantum electrodynamics. 
\end{abstract}

\maketitle

%
%
%
\section{Introduction}
\noindent %
This work is part of a major research effort to use the correlated relativistic energy (and wave function) of few-particle systems as a reference to compute quantum electrodynamical (QED) corrections \cite{MaFeJeMa23,JeFeMa21,JeFeMa22,FeJeMa22b,FeMa23,JeMa23,MaMa24,HoJeMa24,NoMaMa24}. 
Methodologies exist either for computing QED corrections to an uncorrelated but relativistic reference (sometimes called the `$1/Z$' approach \cite{Sh02}), or for computing QED corrections to a correlated but non-relativistic reference state (also called non-relativistic quantum electrodynamics, nrQED approach) \cite{Pa06,KoTs07}. Both directions have several successful applications of high-$Z$~\cite{ArShYePlSo05,VoGlShTuPl14,MaKoSh23,MaKoShTu24} and low-$Z$~\cite{KoHiKa13,KoHiKa14,PaYePa19,PaYeVlPa21,YePaPa22} nuclear charge number systems, respectively. 

In a series of recent work \cite{MaFeJeMa23,JeFeMa21,JeFeMa22,FeJeMa22b,FeMa23,JeMa23,MaMa24,HoJeMa24,NoMaMa24}, we have demonstrated that it should be possible, starting from the Bethe-Salpeter equation and its equal-time time-slice, to define a \emph{relativistic} QED approach, in which a \emph{correlated relativistic} reference state is first (numerically) computed to high precision \cite{JeFeMa21,JeFeMa22,FeJeMa22b,FeMa23}, and then, quantum electrodynamical corrections are computed to it by perturbation theory~\cite{MaFeJeMa23,MaMa24,NoMaMa24}. This second step is still in an early phase of research, and this work is one of the first steps to report actual methodologies and numerical values for the simplest QED corrections to a correlated relativistic reference state. This work considers the simplest, algebraic (energy-independent) correction. When further, important (but more complicated) corrections, \emph{i.e.,} due to retardation, multi-photon, and radiative, self-energy, vertex, and vacuum polarization `effects' will be computed \cite{MaFeJeMa23,MaMa24,NoMaMa24}, then, this direction will complement the already existing $1/Z$ and nrQED approaches, and may help solve the helium puzzle or provide independent tests for (parts of) the other approaches.

Furthermore, the theoretical and methodological developments, based on a correlated relativistic reference state, aim to help extend relativistic quantum chemistry methodologies \cite{DyFaBook07,ReWoBook15,dirac2020} beyond the no-pair approximation.

%
%
%
\section{%
Double-pair ladder correction to the no-pair Dirac-Coulomb(-Breit) energy
\label{sec:doublepair}
} 
The equal-time time-slice of the Bethe-Salpeter equation of two spin-1/2 fermion systems leads to the eigenvalue-like equation, which we name after its pioneers the Salpeter-Sucher equation~\cite{SaBe51,Su58,sucherPhD1958},
\begin{align}
  \left[H_\np+ V_\delta + H_\Delta(E) \right] \Phi = E \Phi \ . \label{eq:SalpeterSucher}
\end{align}
$\Phi(\br_1,\br_2)$ is the equal-time time-slice of the two-particle wave function, depending only on the particles' spatial coordinates, $\br_1$ and $\br_2$. 
$H_\np$ is the no-pair Dirac-Coulomb(-Breit) (DC(B)) Hamiltonian and the $V_\delta + H_\Delta(E)$ term carries the corrections beyond the no-pair processes. 
The `no-pair' term refers to excluding any interaction with the virtual electron–positron pairs, achieved by projecting the two-particle interaction onto the positive-energy subspace \emph{(vide infra)}.
The present work focuses on instantaneous interactions; hence, we drop the energy-dependent $H_\Delta(E)$ term and will work with the algebraic $V_\delta$ contribution.

$H_\np$ is the no-pair Hamiltonian, 
\begin{align}
 H_\np &= h_1 + h_2 + L_{++} V_\mathrm{i} L_{++}\ 
\end{align}
with the
$h_i$ Dirac Hamiltonians for particles $i=1$ and 2,
\begin{align}
  h_i = 
  c \bos{\alpha}_i (-\iim\bos{\nabla}_{\br_i})
  +\beta_i m_i c^2 
  +z_i U_i \; ,
  \label{eq:Dirac}
\end{align}
including the $\bos{\alpha}_i$ and $\beta_i$ Dirac matrices \cite{Di28a}, the $z_i$ electric charge number, and $U_i$ is the scalar potential due to the clamped nuclei. For genuine two-particle, positronium-like systems, $U_i=0$.
The positive-energy subspaces of both $i=1$ and 2 particles are combined to form the $\mcL_{++}$ two-particle positive-energy (no-pair) subspace, for which the projection operator is labelled as $L_{++}$. 
 In practical computations, the energy scale of $h_1$ and $h_2$ is shifted by $-2 m_1c^2$ and $-2m_2c^2$ to match the energy scale of the non-relativistic Hamiltonian, however, we will continue referring to the `positive-energy' and `negative-energy' solutions according to the unshifted Hamiltonians.

The instantaneous interaction energy of particles 1 and 2,
\begin{align}
  V_\ti(\br_1,\br_2)
  = 
  V_\tC (\br_1,\br_2) + V_\tB (\br_1,\br_2) \; ,
\end{align}
is the sum of the Coulomb term, 
\begin{align}
  V_\tC (\br_1,\br_2) 
  = 
  \frac{z_1 z_2}{\left| \br_1-\br_2 \right|} \ , 
\end{align}
and, if instantaneous magnetic effects are included, the Breit term \cite{ReWoBook15,FeJeMa22,FeJeMa22b},
\begin{align}
  V_\tB (\br_1,\br_2) 
  = 
  -z_1 z_2 
  \left[%
    \frac{\balpha_1\cdot\balpha_2}{\left| \br_1-\br_2 \right|}
    +
    \frac{1}{2}\left\lbrace(\balpha_1\cdot\grad_1)(\balpha_2\cdot\grad_2)\left| \br_1-\br_2 \right|\right\rbrace
  \right] \ .
\end{align}

For the double-pair contributions of instantaneous (Coulomb or Coulomb-Breit) interactions, we need to deal with the $V_\delta$ term in Eq.~\eqref{eq:SalpeterSucher} \cite{sucherPhD1958,MaFeJeMa23},
\begin{align}
  V_\delta
  &=
  (L_{++}-L_{--}) V_\ti - L_{++} V_\ti L_{++} \; 
  \nonumber \\
  &=  
  L_{++}V_\ti(1-L_{++}) - L_{--} V_\ti \; 
  \nonumber \\  
  &=
  L_{++}V_\ti L_{--} - L_{--}V_\ti L_{++} - L_{--}V_\ti L_{--} \nonumber \\
  &+
  L_{++}V_\ti (L_{+-}+L_{-+}) - L_{--}V_\ti (L_{+-}+L_{-+}) 
  \ . 
\label{eq:Vdelta}    
\end{align}

First, we solve the no-pair wave equation for the no-pair Dirac-Coulomb(-Breit) Hamiltonian,
\begin{align}
    H_\np \Phi_{\np,i} = E_{\np,i} \Phi_{\np,i} 
    \label{eq:npeq} \ ,
\end{align}
where $i$ is the index of a general state. 
For solutions of the no-pair wave equation, Eq.~\eqref{eq:npeq}, it is convenient to explicitly write out the contributions of $\Phi_{\text{np},i}$ according to the $\mcL_{++}$, $\mcL_\text{BR}=\mcL_{+-}\oplus\mcL_{-+}$, and $\mcL_{--}$ subspaces,
\begin{align}
  \Phi_{\text{np},i}
  =
  \left(%
    \begin{array}{c}
      \phi_{\np,i}^{++} \\[0.10cm]
      \phi_{\np,i}^{\brra} \\
      \phi_{\np,i}^{--}
    \end{array}
  \right) \; ,
\end{align}
in other words, $\phi_{\np,i}^{x}=L_x \Phi_{\text{np},i} L_x$ ($x=++,\text{BR}$ or $--$).
The no-pair Hamiltonian is block diagonal over the $\mcL_{++}, \mcL_\text{BR},$ and $\mcL_{--}$ subspaces.
The non-trivial two-particle solutions, which correspond to the physically relevant bound and excited states, are in the $\mcL_{++}$ positive-energy subspace, hence we can focus on solving the eigenvalue equation on this subspace,
\begin{align}
  H^{++}\phi^{++}_{\np,p}=E^{++}_{\np,p}\phi^{++}_{\np,p} 
  \quad\text{with}\quad 
  H^{++}= {L}_{++}H_\np {L}_{++} \; ,
\end{align}
where $E_{\np,p}^{++} = E_{\np,i}$ with $i\in \mathcal{I}_{++}$ and $\mathcal{I}_{++}$ collects the indexes of the positive-energy solutions of Eq.~\eqref{eq:npeq}.

Over the $\mcL_{--}$ subspace, the no-pair interaction vanishes, and thus, 
\begin{align}
  H^{--}\phi^{--}_{\np,m}=E^{--}_{\np,m}\phi^{--}_{\np,m} 
  \quad\text{with}\quad 
  H^{--}= {L}_{--}H_\np {L}_{--} = {L}_{--} \left( h_1 + h_2 \right) {L}_{--} \; , \label{eq:negenergeq}
\end{align}
where $E_{\np,m}^{--} = E_{\np,i}$ with $i\in \mathcal{I}_{--}$ and $\mathcal{I}_{--}$ collects the indexes of the negative-energy solutions of Eq.~\eqref{eq:npeq}.

The first-order perturbation correction to $E^{++}_{\np,p}$ due to $V_\delta$ is
\begin{align}
   E_p^{(1)}
   &=
   \langle %
     \phi^{++}_{\np,p} | V_\delta | \phi^{++}_{\np,p}
   \rangle 
   = 0 \ ,
   \label{eq:firstorder}
\end{align}
where we used that $\phi^{++}_{\np,p}\in\mcL_{++}$ and $\mcL_{++}$ is orthogonal to $\mcL_{--}$ and the $\mcL_\text{BR}$ subspaces. 
The second-order perturbation theory correction is 
\begin{align}
  E_p^{(2)}
  &= 
  \langle %
    \phi^{++}_{\np,p} %
    | V_\delta \frac{1 - L_{++}}{E^{++}_{\np,p}-H_\np} V_\delta | %
    \phi^{++}_{\np,p} %
  \rangle \ 
  \label{eq:PTtwo}  
  \\
  &=
  \langle %
    \phi^{++}_{\np,p} | 
    (L_{++}V_\ti L_{--} + L_{++}V_\ti L_{+-} + L_{++}V_\ti L_{-+})
    \nonumber \\
  &\quad\quad\quad\quad %
   \frac{1 - L_{++}}{E^{++}_{\np,p}-H_\np}
    (-L_{--}V_\ti L_{++})
    | \phi^{++}_{\np,p} %
  \rangle 
  \\    
  &= 
  -\langle %
    \phi^{++}_{\np,p} %
    | L_{++} V_\ti L_{--} \frac{1-L_{++}}{E^{++}_{\np,p}-H_\np} L_{--} V_\ti L_{++} | %
    \phi^{++}_{\np,p} %
  \rangle \ 
  \\  
  &=
  \langle %
    \phi^{++}_{\np,p} %
    | V_\ti \frac{ L_{--}}{h_1+h_2 - E^{++}_{\np,p}} V_\ti | %
    \phi^{++}_{\np,p} %
  \rangle \ ,
  \label{eq:explicitsecondorder}
\end{align}
where we used the orthogonality of the $\mcL_{++}, \mcL_{--}, \mcL_\text{BR}$ subspaces, and in the last step, $ L_{--}H_\np L_{--} = L_{--}(h_1+h_2)L_{--}$.

As an alternative to the perturbation theory treatment, we can include $V_\delta$ in the Hamiltonian, 
and directly solve the wave equation, 
\begin{align}
  H_\wp  \Phi_{\wp,i} &= E_{\wp,i} \Phi_{\wp,i} \ , \label{eq:wpeq} 
\end{align}
including the `with-pair' Hamiltonian
\begin{align}
  H_\wp =  H_\np + V_\delta = h_1 + h_2 + \left( L_{++} - L_{--} \right)V_\ti  \ , \label{eq:Hwp}
\end{align}
where the contribution of the virtual pairs through the instantaneous interaction is also considered, hence the short name, `with-pair'. 
By solving the with-pair equation, Eq.~\eqref{eq:wpeq}, we obtain a resummation of the instantaneous $V_\delta$ double-pair effects. (It is necessary to note that only the double-pair corrections are summed up, but no crossed-photon contributions are included in this way.)

It is shown in the \som\ that although $H_\wp$ couples the $\mcL_{++}$ and $\mcL_{--}$ subspaces, 
it is block diagonal over $\mcL_{++} \oplus \mcL_{--}$ and $\mcL_{+-} \oplus \mcL_{-+}$, 
which we can exploit during the course of the numerical computations. 
Some further mathematical properties regarding the block-structure of $H_\wp$ are collected in the \som. It is also shown there that although $H_\wp$ in Eq.~\eqref{eq:Hwp} is non-hermitian, it can be rewritten to a hermitian form, and thus, it has real eigenvalues.

\section{Implementation and computational details}
\subsection{Finite basis representation}
For the numerical solution of the no-pair or the with-pair DC(B) wave equation,
\begin{align}
  H^\mathrm{(p)BO} \psi_j^\mathrm{(p)BO} =  E_j^\mathrm{(p)BO} \psi_j^\mathrm{(p)BO} \; ,
\end{align}
we write the wave function as a linear combination of sixteen-dimensional spinors, 
\begin{align}
  \psi^\mathrm{(p)BO}_j
  =
  \sum_{i=1}^{\nb}
  \sum_{q=1}^{16}
    c_{iq,j}
    \mathcal{A}^\mathrm{(p)BO}  \ X^\mathrm{(p)BO} \ 
    {f_{iq}^\mathrm{(p)BO} } \; ,
\end{align}
where `BO' and `pBO' refer to the Born-Oppenheimer \cite{JeFeMa21,JeFeMa22,FeJeMa22,FeJeMa22b} and pre-Born-Oppenheimer \cite{FeMa23} versions of the Ansatz. 
The operator $\mathcal{A}^\mathrm{BO}$ is for the permutational anti-symmetrization of the two electrons
and $\mathcal{A}^\mathrm{pBO}$ is the identity \cite{FeMa23}.

$X^\mathrm{(p)BO}$ is the  restricted kinetic balance matrix \cite{DyFaBook07,ReWoBook15,SuLiKu11,JeFeMa22,FeJeMa22b,FeMa23},
\begin{align}
  X^\mathrm{BO}
  = 
  \text{diag}%
  \left(%
    1\four , 
    \frac{\bsigma_2\four\bp_2}{2m_2c} ,
    \frac{\bsigma_1\four\bp_1}{2m_1c} ,
    \frac{(\bsigma_1\four \bp_1)(\bsigma_2\four \bp_2)}{4m_1m_2c^2} 
  \right) \; , 
  \label{eq:kbBO}
  \\
  X^\mathrm{pBO}
  = 
  \text{diag}%
  \left(%
    1\four , 
    -\frac{\bsigma_2\four\bp}{2m_2c} ,
    \frac{\bsigma_1\four\bp}{2m_1c} ,
    -\frac{(\bsigma_1\four \bp)(\bsigma_2\four \bp)}{4m_1m_2c^2} 
  \right) \; ,
  \label{eq:kbpBO}  
\end{align}
where $\bp=\bp_1-\bp_2$ is the relative momentum of the particles in the pBO case  and the superscript~[4] indicates the matrix dimension. The 1 and 2 subscripts of $\bos{\sigma}^{[4]}$ are the particle indices. They are defined as $\bos{\sigma}_1^{[4]}=\bos{\sigma}^{[2]}\otimes 1^{[2]}$ and $\bos{\sigma}_2^{[4]}=1^{[2]} \otimes \bos{\sigma}^{[2]}$, originating from the block-wise direct product of the single-particle Dirac Hamiltonians in the two-particle theory.
The $X^\mathrm{(p)BO}$ balance was implemented as a metric ($X^{\mathrm{(p)BO}\dagger} X^\mathrm{(p)BO}$), and the Hamiltonian was transformed accordingly ($X^{\mathrm{(p)BO} \dagger} H X^\mathrm{(p)BO}$) \cite{JeFeMa21,JeFeMa22,FeJeMa22,FeJeMa22b,MaFeJeMa23}.

Regarding the kinetic balance, we have considered the inverse kinetic balance or dual kinetic balance conditions \cite{ShTuYePlSo04,SuLiKu11}
at the beginning of this work. 
In the end, good results were obtained with the simple, restricted kinetic balance, Eqs.~\eqref{eq:kbBO}--\eqref{eq:kbpBO}, but it was essential to optimize an (auxiliary) basis set for an appropriate target functional (Sec.~\ref{sec:aux}).

The spinor basis, $f_{iq}^\mathrm{(p)BO}$, is a product of the sixteen-dimensional spinor vector, $\bos{I}_q$, and a spatial function, $\Theta_i(\br)$,
\begin{align}
  f_{iq}^\mathrm{(p)BO} 
  =
  \bos{I}_q \Theta_i^\mathrm{(p)BO}(\br) \ ,
\end{align}
where $\left( \bos{I}_q \right)_p=\delta_{qp}$.
In the BO computations, we used 
explicitly correlated Gaussian (ECG) functions as spatial basis functions \cite{SzJe10,SuVaBook98,MaRe12,Ma19review}, 
\begin{align}
  \Theta_i^\mathrm{BO}(\br) &= \exp \left[- \br^T \underline{\bos{A}}_i \br \right] \ ,
  \label{eq:ecg}
\end{align}
where $\bos{r}$ collects the particles' Cartesian coordinates and $\underline{\bos{A}}_i=\bos{A}_i\otimes I^{[3]}$ is a positive-definite exponent matrix with $\bos{A}_i\in\mathbb{R}^{2\times 2}$. 
In the pBO computations, we have a pseudo-one-particle problem, so the spatial part $\Theta_i^\mathrm{pBO}(\br)$ is a function of the $\br=\br_1-\br_2\in\mathbb{R}^3$ vector. Moreover, for S states, there is no angular dependence, and we chose a simple Gaussian function, similarly to Ref.~\citenum{FeMa23},
\begin{align}
  \Theta^\mathrm{pBO}_i(\br) &= \mathrm{e}^{- a_i \br^2} \ ,
  \label{eq:gauss}
\end{align}
where $a_i\in\mathbb{R}^+$ was a parameter to be optimized.
The non-linear basis parameters, $\bos{A}_i$ (and $a_i$ for pBO), were optimized by minimization of a target functional: (a) the non-relativistic energy (starting parameterization), (b) the no-pair energy (tests and basis set extension), or (c) the perturbative double-pair (Coulomb) correction.  

To construct the matrix representation of $H_\np$ or $H_\wp$, it was necessary to compute the $L_{++}$ or the $L_{++}$ and $L_{--}$ projector(s) over the basis space. The initially used cutting projector \cite{JeFeMa22}, as well as the most recently developed \honehtwo\ projector \cite{HoJeMa24} were tested in the present work. 
The \honehtwo\ projection scheme automatically performs a rigorous selection of the $\mcL_{++}/\mcL_{+-}/\mcL_{-+}/\mcL_{--}$ two-particle subspaces based on the consecutive diagonalization of the one-particle Hamiltonians over the two-particle basis space. 

In former no-pair relativistic ECG computations, the cutting projector was found to be efficient, although it is necessarily approximate (for conceptual reasons), as it identifies (approximates) the $\mcL_{++}$ subspace with the subspace spanned by all eigenstates of the non-interacting two-particle Hamiltonian, $h_1+h_2$, that have an energy larger than a pre-defined $E_\text{th}$ threshold energy. Similarly, the cutting projector approximation of $\mcL_{--}$ is obtained as all non-interacting two-particle states with energy less than $-2(m_1+m_2)c^2$. By this (simple) construction, both the $\mcL_{++}$ and the $\mcL_{--}$ cutting subspaces have some Brown-Ravenhall (BR) contamination. The \honehtwo\ projection approach \cite{HoJeMa24} eliminates this problem, but it is technically (and numerically) more subtle than the simple energy-cutting scheme. So far, we have seen only a tiny effect of the cutting approach' BR contamination in variational ECG computations of the no-pair energy; the difference was on the order of the finite-basis convergence error \cite{JeFeMa22,FeJeMa22b,HoJeMa24}. In all tests carried out in the present work for the double-negative pair corrections, we have not found any difference larger than the finite-basis set error. So, the two different projection approaches (and any numerical differences) are not discussed further in this work.

\subsection{Variational optimization of the basis functions}
High-precision energies with an ECG (Gaussian) basis set can be obtained if a suitable target functional is defined and subsequently minimized (or maximized) with respect to the non-linear basis parameters.
Most often, the non-relativistic energy is chosen as the target functional, which is bounded from below, and it can be minimized in numerical procedures \cite{SuVaBook98}. 
In the relativistic case, the no-pair DC(B) energy can likewise be minimized to optimize the basis parameters, since it is also bounded from below. 
Of course, the no-pair DC(B) energy minimization is more computationally expensive than the non-relativistic energy minimization, due to the recomputation of the $L_{++}$ projector for every new trial parameter set.
We also note that for small nuclear charge numbers, minimizing the non-relativistic energy already yields a sufficiently accurate basis parameterization for the no-pair DC(B) energy \cite{JeFeMa21,JeFeMa22,FeJeMa22b,MaFeJeMa23,NoMaMa24,HoJeMa24}, even when high precision is required.

However, the basis sets that deliver us accurate no-pair (or non-relativistic) energies do not necessarily provide a reliable description of the $\mcL_{--}$ subspace required for the double-pair corrections. 
Moreover, since the with-pair Hamiltonian, Eq.~\eqref{eq:Hwp}, is not bounded from below, the basis representation of the $\mcL_{--}$ subspace cannot be systematically improved through the minimization of the with-pair energy.
In the next subsection, we first write the working formulae for the second-order perturbation-theory (PT) corrections, and then, identify the quantity for which an auxiliary basis set is optimized to improve the $\mcL_{--}$ representation for the double-pair computations.

\subsection{Spectral representation of the perturbative corrections}
The second-order PT correction, Eq.~\eqref{eq:explicitsecondorder}, for $V_\ti=V_{\rm C}+V_{\rm B}$ is written as
\begin{align}
  E^{(2)}
  &= 
  E_{\text{CC}}^{(2)} + 2E_{\text{CB}}^{(2)} + E_{\text{BB}}^{(2)} \; . %
  \label{eq:E2spectral} 
\end{align}
Then, we insert the spectral representation (truncated according to the finite basis set) to obtain a working formula for $E_{\text{CC}}^{(2)}$,
\begin{align}
  E_{\text{CC}}^{(2)} 
  &= 
  \sum_{m=1}^{N}
    \frac{%
      \langle \phi^{++}_{{\np}}|V_\mathrm{C}|\phi^{--}_{0,m} \rangle
      \langle \phi^{--}_{0,m}|V_\mathrm{C} |\phi^{++}_{\np} \rangle
    }{E^{--}_{0,m}-E^{++}_{\np} } \ ,
    \label{eq:E2spectralCC} 
\end{align}
where 
$E^{--}_{0,m}=E^{--}_{\np,m}$ ($\phi^{--}_{0,m}=\phi^{--}_{\np,m}$) is the eigenvalue (eigenfunction) of the non-interacting Hamiltonian, $h_1+h_2$, over the $\mcL_{--}$ subspace. 
The Breit-Breit and Coulomb-Breit interaction corrections were computed similarly,
\begin{align}
  E_{\mathrm{BB}}^{(2)} 
  &= 
  \sum_{m=1}^{N}
    \frac{%
      \langle \phi^{++}_{\np}|V_\mathrm{B} |\phi^{--}_{0,m} \rangle
      \langle \phi^{--}_{0,m}|V_\mathrm{B} |\phi^{++}_{\np} \rangle
    }{%
      E^{--}_{0,m} - E^{++}_{\np}
    }  \ , \label{eq:E2spectralBB} 
    \\
  E_{\mathrm{CB}}^{(2)} 
  &= 
  \frac{1}{2}
  \sum_{m=1}^{N}  
  \left[%
      \frac{%
        \langle \phi^{++}_{\np}|V_\mathrm{C}|\phi^{--}_{0,m} \rangle
        \langle \phi^{--}_{0,m}|V_\mathrm{B} |\phi^{++}_{\np} \rangle 
      }{%
        E^{--}_{0,m} - E^{++}_{\np} 
      } 
      + 
      \frac{%
        \langle \phi^{++}_{\np}|V_\mathrm{B}|\phi^{--}_{0,m} \rangle
        \langle \phi^{--}_{0,m}|V_\mathrm{C} |\phi^{++}_{\np} \rangle 
      }{%
        E^{--}_{0,m} - E^{++}_{\np} 
      } 
  \right] \nonumber \\
  &=
  \text{Re}
  \left[%
  \sum_{m=1}^{N}  
      \frac{%
        \langle \phi^{++}_{\np}|V_\mathrm{C}|\phi^{--}_{0,m} \rangle
        \langle \phi^{--}_{0,m}|V_\mathrm{B} |\phi^{++}_{\np} \rangle 
      }{%
        E^{--}_{0,m} - E^{++}_{\np} 
      } 
  \right] \; .
  \label{eq:E2spectralCB}
\end{align}

\subsection{Auxiliary basis set optimization \label{sec:aux}}
As an appropriate target functional to improve the finite-basis representation of the $\mcL_{--}$ subspace, we used the double-pair Coulomb correction, Eq.~\eqref{eq:E2spectralCC}.
$E_{\mathrm{CC}}^{(2)}$ is negative, since the numerator of Eq.~\eqref{eq:E2spectralCC} is positive, $|\langle \phi^{++}_{\np} | V_\mathrm{C} | \phi^{--}_{0,m}\rangle|^2>0$, and the denominator is negative, $E^{--}_{0,m}-E^{++}_{\np}<0$ (for all physically relevant $E^{++}_{\np}$ energies on $\mcL_{++}$).
An (auxiliary) basis set for the finite basis representation of the $\mcL_{--}$ subspace was optimized by minimizing $E_{{\rm CC}}^{(2)}$.

There are some subtleties regarding the practical realization of this optimization procedure. The $L^\text{aux}_{--}$ and $L^\text{ref}_{++}$ projectors, constructed over the finite-dimensional (and different) auxiliary (aux) and reference (ref) basis sets are not strictly orthogonal (the reference basis set is the basis set of the $\phi^{++}_\text{np}$ reference state). 
A numerically stable minimization of $E_\text{CC}^{(2)}$ was possible only when the orthogonality of $\mcL_{--}^\text{aux}$ was enforced (by Gram-Schmidt) orthogonalization of the auxiliary basis vectors spanning $\mcL_{--}^\text{aux}$ with respect to the reference basis states spanning the $\mcL_{++}^\text{ref}$ and $\mcL_{+-/-+}^\text{ref}$ subspaces. 
During the auxiliary basis optimization stage, $N=N_\text{aux}$ in Eq.~\eqref{eq:E2spectralCC}. 
After the optimization of the auxiliary basis set is completed, the auxiliary ($N_\text{aux}$) and the reference ($\nb$) basis sets were merged, and both the $\phi_\np^{++}$ reference state and the second-order corrections, Eq.~\eqref{eq:E2spectralCC}, were computed with $N=N_\text{aux}+\nb$ basis functions.
In this merged basis, the $\mcL_{--}$, $\mcL_{+-/-+}$, and $\mcL_{++}$ subspaces are orthogonal by construction, 
hence, either the perturbative calculations, Eq.~\eqref{eq:E2spectral}, or the finite-basis solution of the with-pair equation, Eq.~\eqref{eq:wpeq}, could be carried out without additional orthogonalization steps. 

We emphasize that the auxiliary basis optimization was performed with the $V_\text{C}$ (Coulomb) interaction. We have not used the $V_\text{BB}$ (Breit-Breit) interaction for this purpose for the excessively large coupling of the (instantaneous) Breit interaction with the $\mcL_{--}$ subspace \emph{(vide infra)}.
The optimization of $E_\text{CC}^{(2)}$ within the sixteen-component framework, involving the repeated construction of the projectors, was computationally expensive. For this reason, and particularly during the development stage, we have extensively studied genuine two-particle systems (Ps, H, Mu, $\mu$H), for which the spatial basis set was small (pseudo-one-particle, sixteen-component representation).

\section{Numerical results and comparison with nrQED values}
The newly computed double-pair corrections to the no-pair reference and the finite-basis solution of the with-pair eigenvalue equation were extensively tested with well-established nrQED values through the $\alpha$ fine-structure dependence of our relativistic QED computations.

\subsection{Comparison with nrQED through the $\alpha$ expansion}
The perturbative energy corrections, Eq.~\eqref{eq:E2spectral}, and the with-pair energy (non-trivially) depend on the value of the $\alpha$ fine-structure constant,
\begin{align}
  E^{(2)}(\alpha) = \epsi_{3} \alpha^3 + \epsi_{4,0} \alpha^4 + \epsi_{4,1} \alpha^4 \ln\alpha  + \ldots
  \label{eq:Etwoalpha}
\end{align}
and
\begin{align}
  E_{\text{wp}}(\alpha) = \epsi_{0} + \epsi_{2} \alpha^2 + \epsi_{3} \alpha^3 + \epsi_{4,0} \alpha^4 + \epsi_{4,1} \alpha^4 \ln\alpha +\ldots \;,
  \label{eq:Ewpalpha}
\end{align}
where the $\epsi_{k}$ coefficients were determined numerically by fitting these functional forms to the energy values computed for a series of slightly different $\alpha$ values. Then, the fitted $\epsi_{k}$ coefficients can be directly compared with the $\epsilon_{k}$ nrQED values, where they are available (the different $\epsi$ and $\epsilon$ symbols are used to distinguish the two approaches).

We emphasize that our primary computational results are $E^{(2)}$ and $E_{\wp}$, on the left of Eqs.~\eqref{eq:Etwoalpha}  and \eqref{eq:Ewpalpha}. At the same time, the nrQED approach delivers us the coefficients corresponding to the right-hand side of the equations.

\subsubsection{nrQED expressions \label{sec:nrqed}}
It was discussed in earlier work \cite{JeFeMa22,FeJeMa22,FeJeMa22b} that 
the $\alpha$ dependence of the no-pair energy was in excellent agreement with nrQED using the
$\epsilon_0$ non-relativistic energy, the $\epsilon_2$ leading-order relativistic energy (corresponding to the DC or DCB Hamiltonians), and the $\alpha^3$-order $\epsilon_{3}$ for the positive-energy ($++$) ladder corrections of the Coulomb (and Coulomb-Breit) photon exchanges~\cite{FuMa54,sucherPhD1958}. Even the $\alpha^4\ln\alpha$-order $\epsilon_{4,1}$ logarithmic Coulomb-photon correction \cite{DrKhMiYe93,KhMiYe93,Zh96} could have been identified in the $\alpha$-dependence of the high-precision no-pair energy of pBO \cite{FeMa23} and BO atomic systems~\cite{NoMaMa24}.

\paragraph{Pre-BO systems}
The sum of the no-pair and the double-pair $\alpha^3\Eh$-order Coulomb photon corrections is, 
according to Eq.~(3.7) of Ref.~\citenum{FuMa54},
\begin{align}
  \epsilon^{++,--}_{3,\text{CC}}(m_1,m_2) 
  = 
  -\frac{2}{3} \frac{\mu^3}{\pi} \left( \frac{2}{m_1^2} + \frac{1}{m_1 m_2} +  \frac{2}{m_2^2} \right) \; ,
  \label{eq:pBOCCwp}
\end{align}
where $\mu=(m_1^{-1}+m_2^{-1})^{-1}$ is the reduced mass.

As to the double-pair Coulomb photon correction, the $\epsilon_{3,{\text{CC}}}^{--}$ nrQED value (written separately from the no-pair $\epsilon_{3,{\text{CC}}}^{++}$ term) can be (numerically) calculated for the genuine (pBO) two-particle systems by using the integral formula of Fulton and Martin, \emph{i.e.,} the two-pair contributions in the Eqs. (3.1a)–(3.6) of Ref.~\citenum{FuMa54} (we note the missing $1/(k^2 E_1 E_2)$ factor in the integrand of Eq.~(S12) of Ref.~\citenum{FeMa23}),
\begin{align}
  \epsilon_{3,{\text{CC}}}^{--}(m_1,m_2)
  =
  -\frac{2\mu^3}{\pi} \int_0^\infty \text{d} k \, \frac{1}{k^2 E_1 E_2} \, \frac{(E_1 - m_1)(E_2-m_2)}{E_1+E_2+m_1+m_2}  
  \label{eq:CCmm}
\end{align}
with $E_a=\sqrt{m_a^2+k^2}$ ($a=1,2$). 
For the special case of Ps (and two-electron BO systems), $m_1=m_2=1$ (in units of $m_\text{e}$), the integral can be calculated analytically, and it is
\begin{align}
  \epsilon_{3,{\text{CC}}}^{--}(1,1)
  =
  -\frac{1}{8\pi}\left(\frac{5}{3}-\frac{\pi}{2}\right) \ .
  \label{eq:CC11}
\end{align}
Furthermore, the $\alpha^4\ln\alpha\Eh$ term for the Dirac-Coulomb case is also available from the literature \cite{KhMiYe93,FeMa23}
\begin{align}
  \epsilon_{4,1,{\text{CC}}}^{--}(1,1)
  =
  -\frac{1}{16} \ .
  \label{eq:CC11}
\end{align}
As outlined in the \som, we could obtain an analytic expression for the Coulomb-Breit double-pair corrections (along similar lines to the CC$--$ calculation of Ref.~\citenum{sucherPhD1958}), 
\begin{align}
  \epsilon_{3,{\text{CB}}}^{--}(1,1)
  =
  -\frac{1}{8\pi}\left(\pi-2\right) \ .
  \label{eq:CBmm}  
\end{align}

\paragraph{BO systems}
For the BO systems, the $\alpha^3\Eh$-order CC$--$ nrQED expression is, according to the Eq.~(4.26b) of Ref.~\citenum{sucherPhD1958}, 
\begin{align}
  \epsilon_{3,\mathrm{CC}}^{--} 
  &= 
  -\left( \frac{5}{3}-\frac{\pi}{2}\right) 
  \langle \delta(\br_{12}) \rangle_\mathrm{nr}
  \label{eq:epcCCmmBO} \ .  
\end{align}
We note that this expression is analogous to Eq.~\eqref{eq:CC11}, but instead of using the analytic result (with $m=m_1=m_2=1$) for 
$\langle \delta(\br_{12})\rangle_\text{nr}=\frac{1}{8\pi}$ 
($n=1,l=0$, ground state),
we compute $\langle \delta(\br_{12})\rangle_\text{nr}$ numerically.

The CB$--$ nrQED expression for general BO systems is (\som) 
\begin{align}
  \epsilon_{3,\mathrm{CB}}^{--} 
  &= 
  -\left( \pi-2\right) 
  \langle \delta(\br_{12}) \rangle_\mathrm{nr} 
  \label{eq:epcCBmmBO} \ .
\end{align}
For completeness, we also reiterate the $\alpha^3\Eh$-order no-pair correction for general BO systems, 
Eq.~(3.99) and Eq.~(5.64) of \cite{sucherPhD1958}, 
\begin{align}
  \epsilon_{3,\mathrm{CC}}^{++} 
  &=
  -\left(\frac{\pi}{2}+\frac{5}{3}\right) 
  \langle \delta(\br_{12})\rangle_\text{nr}
  \nonumber \\
  \epsilon_{3,\mathrm{CB}}^{++} 
  &= 
  4\left( \frac{\pi}{2}+1 \right) 
  \langle \delta(\br_{12})\rangle_\text{nr} \; .
\end{align}

The $\alpha^4\Eh$-order $\epsilon_{4,0}$ nrQED expression for the no-pair or the no-pair plus double-pair energies 
cannot be written separately from several other nrQED terms entering at this order (due to divergences that are cancelled only in the final result) \cite{Pa06}. 
At the same time, the $\alpha^4\ln\alpha\Eh$-order $\epsilon_{4,1}$ contribution is known for the Coulomb part of the problem \cite{KhMiYe93,Zh96}
\begin{align}
  \epsilon_{4,1,{\rm CC}}^{++} 
  = 
  \frac{\pi}{2} %
  \langle \delta(\br_{12}) \rangle_\text{nr} \ . 
\end{align}

\subsubsection{Discussion of the numerical results}
Tables \ref{tab:pBO} and \ref{tab:BO} show the convergence of the second-order perturbation-theory (PT) energies, Eqs.~\eqref{eq:E2spectral}--\eqref{eq:E2spectralCB}, for the double-pair instantaneous interaction corrections.
The tables also show the $\epsi_3$ values obtained from repeating the computations for a series of $\alpha$ values and fitting the $\alpha$-dependent function of Eq.~\eqref{eq:Etwoalpha} to the data. 
These $\epsi_3$ values (free of higher-order $\alpha$ contributions) can be directly compared with the $\epsilon_3$ double-pair correction nrQED values, Eqs.~\eqref{eq:CCmm}--\eqref{eq:CBmm} and \eqref{eq:epcCCmmBO}--\eqref{eq:epcCBmmBO}.

Good convergence of the corrections was obtained only if the reference-state basis set was extended with auxiliary functions, optimized by minimization of $E_\text{CC}^{(2)}$. In some cases, without an optimized auxiliary basis set, even the order or magnitude of the correction was wrong, indicating an incomplete representation of $\mcL_{--}$. 
The value of the correction and its convergence were checked by extracting the $\epsi_3$ $\alpha^3$-order coefficient of the corrections, and comparing it with the $\epsilon_3$ nrQED value. The convergence of the correction (ref+aux) is very good and the deviation from the nrQED value is comparable to the convergence of the (non-relativistic or) no-pair energy of the reference state (Tables~S2 and S3, \som). 
All computations were carried out in quadruple precision arithmetic.

The double-pair Breit (BB$--$) corrections are also shown in Tables~\ref{tab:pBO} and \ref{tab:BO}, although they are unphysically large (and have a too strong high-$\alpha$-order dependence). This feature can be understood by the (unphysically) strong coupling of the Breit interaction of the positive- and negative-energy subspaces. This excessive coupling is expected to be attenuated when the complete (non-instantaneous) transverse interaction is considered, which will be the subject of future work. 

Using the (ref+aux) basis sets, the with-pair Dirac-Coulomb equation was solved for all systems studied in this work. The convergence of the with-pair energy for selected basis sizes is also shown in Tables~S2 and S3.
We note that the with-pair Hamiltonian is not bounded from below, so the energy convergence is not necessarily monotonic with the basis size.

For the largest basis sets, the with-pair computation was repeated with slightly different $\alpha$ values, and we fitted the $\epsi_k$ coefficients of Eq.~\eqref{eq:Ewpalpha} to this dataset. Table~\ref{tab:wpDC} shows the $\epsi_k$ fitted values and the difference of the relevant term (in $\Eh$) from the nrQED value.  
Regarding the nrQED terms, we had to consider the same terms as in Refs.~\cite{JeFeMa22,FeJeMa22b,NoMaMa24} (reiterated in Sec.~\eqref{sec:nrqed} for completeness), appended with the CC$--$ terms, Eqs.~\eqref{eq:CCmm} and \eqref{eq:epcCCmmBO}.
The deviation of the nrQED and fitted contributions to the energy (in $\Eh$) was found to be approximately constant for the fitted terms, suggesting the correctness of the coefficients in Table~\ref{tab:wpDC}. (The larger deviation of some coefficients of H and $\mu$H in Table~\ref{tab:pBO} can be explained by the smallness of these contributions and the convergence of the reference energy.)

\section{Conclusions}
Double-pair corrections with instantaneous interactions are computed to a no-pair Dirac-Coulomb(-Breit) (DC(B)) reference state of two-spin-1/2-fermion systems. 
The theoretical framework is based on the equal-time Bethe-Salpeter equation. In practice, first, the no-pair DC(B) equation was solved over a finite basis set of explicitly correlated (Gaussian) functions, and then, perturbative instantaneous double-pair corrections were computed. For well-converged pair-corrections, it was essential to optimize an auxiliary basis set for better representing the $\mcL_{--}$ double-pair subspace in addition to the basis representation optimized by minimization of the no-pair energy. Nevertheless, the simplest `restricted' kinetic balance condition was sufficient for converging the results, if the reference basis set was extended with the auxiliary basis functions optimized to the double-pair Coulomb correction.

Numerical results were reported for genuine two-particle systems (Ps, Mu, H, and $\mu$H) and for the He isoelectronic series with $Z=2,3,4$ nuclear charge numbers (He~1S$_0$, He~2S$_0$, Li$^+$~1S$_0$, and Be$^{2+}$~1S$_0$). In future work, the computations can be extended to small molecular systems.
The $\alpha$ fine-structure dependence of the perturbative corrections was obtained in excellent agreement with the available non-relativistic quantum electrodynamics (nrQED) corrections at the relevant $\alpha$ order. 
In addition to a perturbative treatment, we have also considered a `with-pair' equation in which the double-pair instantaneous corrections (an algebraic term) increment the no-pair Dirac-Coulomb Hamiltonian. The numerical solution of this with-pair equation, in principle, delivers resummation in the double-pair terms. The $\alpha$ dependence of the with-pair energies was obtained in excellent agreement with the perturbative and nrQED results. However, for the smallness of the double-pair Coulomb correction, it remains sufficient to continue with the perturbative computation. There is no technical difficulty in including the Breit term in addition to the Coulomb interaction, but for physically relevant double-pair corrections, we aim to develop a perturbative treatment for the double-pair transverse (instead of the Breit) corrections in the future. This requires accounting for beyond instantaneous kernels which appear in $H_\Delta(E)$ of Eq.~\eqref{eq:SalpeterSucher}. Non-radiative and radiative contributions are currently explored in the research group along the lines initiated in Refs.~\cite{MaFeJeMa23,MaMa24,NoMaMa24}.

\begin{table}[h]
  \caption{%
    Perturbative double-pair energy correction, $E^{(2)}$ (for Coulomb and Breit photon exchanges), 
    to the no-pair Dirac-Coulomb energy for ground states of genuine two-spin-1/2 systems.
    For the largest basis sets, the $\alpha^3\Eh$-order coefficient ($\epsi_3$) and the nrQED value ($\epsilon_3$) are also shown.
    \label{tab:pBO}
  }  
\scalebox{0.95}{%
  \begin{tabular}{@{}l@{}c@{} rrrr c rrrr c rrrr@{}}
    \hline\hline\\[-0.35cm] 
    $\nb$ / $\naux$~$^\text{a}$ &&
    \multicolumn{1}{c}{0} & \multicolumn{1}{c}{10} & \multicolumn{1}{c}{20} &  \multicolumn{1}{c}{30} &&
    \multicolumn{1}{c}{0} & \multicolumn{1}{c}{10} & \multicolumn{1}{c}{20} &  \multicolumn{1}{c}{30} && 
    \multicolumn{1}{c}{0} & \multicolumn{1}{c}{10} & \multicolumn{1}{c}{20} &  \multicolumn{1}{c}{30}   \\
    \cline{1-6} \cline{8-11} \cline{13-16} \\[-0.35cm]
\multicolumn{1}{r}{\scalebox{0.9}{$\text{Ps}=\lbrace\text{e}^-,\text{e}^+\rbrace$}} &&
    \multicolumn{4}{c}{$E_\mathrm{CC}^{(2)}$ [n$E_\mathrm{h}$]} &&
    \multicolumn{4}{c}{$E_\mathrm{CB}^{(2)}$ [n$E_\mathrm{h}$]} && 
    \multicolumn{4}{c}{$E_\mathrm{BB}^{(2)}$ [$\mu E_\mathrm{h}$]} \\ 
    \cline{3-6} \cline{8-11} \cline{13-16} \\[-0.35cm]
    30 && --1.465 & --1.466 & --1.476 & --1.476  && --17.47 & --17.47 & --17.49 & --17.49 && --26.42 & --26.42 & --26.42 & --26.42 \\
    40 && --1.474 & --1.474 & --1.476 & --1.476  && --17.49 & --17.49 & --17.49 & --17.49 && --26.45 & --26.45 & --26.45 & --26.45  \\
    50 && --1.476 & --1.476 & --1.476 & --1.477  && --17.49 & --17.49 & --17.49 & --17.49 && --26.46 & --26.46 & --26.46 & --26.46 \\
    \multicolumn{1}{r}{$\epsi_{3}$$^\text{b}$} &
      & & & & --1.476 && & & & --17.64 && & & & --87 \hspace{0.35cm}  \\
    \multicolumn{1}{r}{$\epsilon_{3}$$^\text{c}$} &%
      & & & & --1.483 && & & & --17.65 && & & & \multicolumn{1}{c}{---} \\
    \hline\\[-0.35cm]
\multicolumn{1}{r}{\scalebox{0.9}{$\text{Mu}=\lbrace\text{e}^-,\mu^+\rbrace$$^\text{d}$}} && 
    \multicolumn{4}{c}{$E_\mathrm{CC}^{(2)}$ [p$E_\mathrm{h}$]} && 
    \multicolumn{4}{c}{$E_\mathrm{CB}^{(2)}$ [p$E_\mathrm{h}$]} && 
    \multicolumn{4}{c}{$E_\mathrm{BB}^{(2)}$ [$\mu E_\mathrm{h}$]} \\ 
    \cline{3-6} \cline{8-11} \cline{13-16} \\[-0.35cm]
    30 && --0.13 & --1.23 & --1.51 & --1.41  && --22.0 & --31.7 & --30.7 & --31.5 && --1.01 & --1.01 & --1.01 & --1.01 \\
    40 && --0.25 & --1.21 & --1.50 & --1.40  && --25.4 & --30.2 & --30.7 & --31.2 && --1.01 & --1.01 & --1.01 & --1.01  \\
    50 && --0.34 & --1.12 & --1.46 & --1.38  && --26.4 & --31.7 & --31.0 & --31.5 && --1.01 & --1.01 & --1.01 & --1.01 \\
    \multicolumn{1}{r}{$\epsi_{3}$$^\text{b}$} & 
    & & & & --1.46 && & & & --30.2 && & & & --3.3 \hspace{0.06cm}  \\
    \multicolumn{1}{r}{$\epsilon_{3}$$^\text{c}$} & 
    & & & & --1.39 && & & & \multicolumn{1}{c}{---} && & & & \multicolumn{1}{c}{---} \\
    \hline\\[-0.35cm]
\multicolumn{1}{r}{\scalebox{0.9}{$\text{H}=\lbrace\text{e}^-,\text{p}^+\rbrace$$^\text{d}$}} &&  
    \multicolumn{4}{c}{$E_\mathrm{CC}^{(2)}$ [f$E_\mathrm{h}$]} && 
    \multicolumn{4}{c}{$E_\mathrm{CB}^{(2)}$ [f$E_\mathrm{h}$]} && 
    \multicolumn{4}{c}{$E_\mathrm{BB}^{(2)}$ [n$E_\mathrm{h}$]} \\ 
    \cline{3-6} \cline{8-11} \cline{13-16} \\[-0.35cm]
    30 && --0.193 & --18.2 & --22.0 & --21.9  && --287 & --285 & --369 & --381 && --11.5 & --11.5 & --11.5 & --11.5 \\
    40 && --0.391 & --28.8 & --29.1 & --29.1  && --335 & --468 & --460 & --461 && --11.5 & --11.5 & --11.5 & --11.5  \\
    50 && --0.564 & --28.6 & --28.9 & --28.8  && --360 & --472 & --466 & --467 && --11.5 & --11.5 & --11.5 & --11.5 \\
    \multicolumn{1}{r}{$\epsi_{3}$$^\text{b}$} &
    & & & & --15 \hspace{0.14cm}  && & & & --472 && & & & --380 \hspace{0.01cm} \\
    \multicolumn{1}{r}{$\epsilon_{3}$$^\text{c}$} &
    & & & & --18.2 && & & & \multicolumn{1}{c}{---} && & & & \multicolumn{1}{c}{---} \\
    \hline\\[-0.35cm]
\multicolumn{1}{r}{\scalebox{0.9}{$\mu\text{H}=\lbrace \mu^-,\text{p}^+\rbrace$$^\text{d}$}} &&  
    \multicolumn{4}{c}{$E_\mathrm{CC}^{(2)}$ [n$E_\mathrm{h}$]} && 
    \multicolumn{4}{c}{$E_\mathrm{CB}^{(2)}$ [$\mu E_\mathrm{h}$]} && 
    \multicolumn{4}{c}{$E_\mathrm{BB}^{(2)}$ [m$E_\mathrm{h}$]} \\ 
    \cline{3-6} \cline{8-11} \cline{13-16} \\[-0.35cm]
    30 && --70.1 & --70.4 & --80.9 & --81.2  && --1.14 & --1.14 & --1.16 & --1.16 && --3.58 & --3.58 & --3.58 & --3.58 \\
    50 && --80.0 & --80.0 & --81.7 & --81.7  && --1.16 & --1.16 & --1.16 & --1.16 && --3.58 & --3.58 & --3.58 & --3.58 \\
    60 && --80.0 & --80.0 & --81.7 & --81.7  && --1.16 & --1.16 & --1.16 & --1.16 && --3.58 & --3.58 & --3.58 & --3.58 \\
    \multicolumn{1}{r}{$\epsi_{3}$$^\text{b}$} &
    & & & & --78.9 && & & & --1.16 && & & & --11.7  \\
    \multicolumn{1}{r}{$\epsilon_{3}$$^\text{c}$} &
    & & & & --81.9 && & & & \multicolumn{1}{c}{---} && & & & \multicolumn{1}{c}{---} \\
    \hline\hline
  \end{tabular}
}
  \begin{flushleft}
    $^\text{a}$ %
      $\nb$ is the number of the basis functions optimized for (the non-relativistic energy of) the reference state and $N_\mathrm{aux}$ is the auxiliary basis set optimized for $E^{(2)}_\text{CC}$, Eq.~\eqref{eq:E2spectralCC}. The listed corrections were computed by merging the reference and auxiliary basis sets. \\
    $^\text{b}$ %
      $\epsi_3$ is obtained from $\alpha$ scaling and fitting the energy correction according to Eq.~\eqref{eq:Etwoalpha}. \\
    $^\text{c}$ %
      $\epsilon_3$ is the $\alpha^3\Eh$-order nrQED value, Eqs.~\eqref{eq:CCmm}--\eqref{eq:CBmm}. \\
    $^\text{d}$ %
      $m_\mu= 206.768\ 283\ 0 \, m_\text{e}$, 
      $m_\text{p}= 1836.152\ 673\ 425\ 726  \, m_\text{e}$ \cite{MoNeTaTi25}.
  \end{flushleft}    
\end{table}

\begin{table}[h]
  \caption{%
    Perturbative double-pair energy correction, $E^{(2)}$ (for Coulomb and Breit photon exchanges), 
    to the no-pair Dirac-Coulomb energy of two-electron atomic systems. 
    For the largest basis sets, the $\alpha^3\Eh$-order dependence ($\epsi_3$) and the nrQED value ($\epsilon_3$) are also shown.   
    \label{tab:BO}
  }  
\scalebox{0.97}{%
  \begin{tabular}{@{}l@{}c@{} rrrr c rrrr c rrrr@{}}
    \hline\hline\\[-0.35cm] 
    $\nb$ / $\naux$~$^\text{a}$ &&
    \multicolumn{1}{c}{0} & \multicolumn{1}{c}{20} & \multicolumn{1}{c}{50} &  \multicolumn{1}{c}{100} &&
    \multicolumn{1}{c}{0} & \multicolumn{1}{c}{20} & \multicolumn{1}{c}{50} &  \multicolumn{1}{c}{100} && 
    \multicolumn{1}{c}{0} & \multicolumn{1}{c}{20} & \multicolumn{1}{c}{50} &  \multicolumn{1}{c}{100}   \\
    \cline{1-16} \\[-0.35cm]
\multicolumn{1}{r}{\scalebox{1.}{$\text{He}$~(1S$_0$)}} &&
    \multicolumn{4}{c}{$E_\mathrm{CC}^{(2)}$ [n$E_\mathrm{h}$]} && 
    \multicolumn{4}{c}{$E_\mathrm{CB}^{(2)}$ [n$E_\mathrm{h}$]} && 
    \multicolumn{4}{c}{$E_\mathrm{BB}^{(2)}$ [$\mu E_\mathrm{h}$]} \\
    \cline{3-6} \cline{8-11} \cline{13-16} \\[-0.35cm]
     500 && --3.928 & --3.930 & --3.930 & --3.933  && --46.04 & --46.26 & --46.26 & --46.26 && --35.19 & --35.19 & --35.19 & --35.19 \\
     750 && --3.949 & --3.950 & --3.950 & --3.951  && --46.30 & --46.30 & --46.30 & --46.30 && --35.27 & --35.27 & --35.27 & --35.27  \\
    1000 && --3.956 & --3.956 & --3.956 & --3.956  && --46.31 & --46.31 & --46.31 & --46.31 && --35.34 & --35.34 & --35.34 & --35.34 \\
    \multicolumn{1}{r}{$\epsi_{3}$$^\text{b}$} & 
     & & & & --3.964 && & & & --47.15 && & & & --86 \hspace{0.35cm}  \\
    \multicolumn{1}{r}{$\epsilon_{3}$$^\text{c}$} & 
     & & & & --3.962 && & & & --47.18 && & & & \multicolumn{1}{c}{---} \\
    \hline\\[-0.35cm]
\multicolumn{1}{r}{\scalebox{1.}{$\text{He}$~(2S$_0$)}} &&
    \multicolumn{4}{c}{$E_\mathrm{CC}^{(2)}$ [n$E_\mathrm{h}$]} && 
    \multicolumn{4}{c}{$E_\mathrm{CB}^{(2)}$ [n$E_\mathrm{h}$]} && 
    \multicolumn{4}{c}{$E_\mathrm{BB}^{(2)}$ [$\mu E_\mathrm{h}$]} \\ 
    \cline{3-6} \cline{8-11} \cline{13-16} \\[-0.35cm]
    200 && --0.292 & --0.294 & --0.292 & --0.309 && --4.66 & --4.67 & --4.66 & --4.66 && --2.82 & --2.83 & --2.86 & --2.87 \\
    300 && --0.292 & --0.294 & --0.293 & --0.309 && --4.68 & --4.69 & --4.69 & --4.71 && --3.01 & --3.01 & --3.01 & --3.02 \\
    400 && --0.295 & --0.296 & --0.294 & --0.310 && --4.70 & --4.70 & --4.70 & --4.73 && --3.02 & --3.02 & --3.03 & --3.03 \\
    \multicolumn{1}{r}{$\epsi_{3}$$^\text{b}$} & 
    & & & & --0.313 && & & & --4.32 && & & & --2.86   \\
    \multicolumn{1}{r}{$\epsilon_{3}$$^\text{c}$} & 
    & & & & --0.322 && & & & --4.34 && & & & \multicolumn{1}{c}{---} \\
    \hline\\[-0.35cm]    
\multicolumn{1}{r}{\scalebox{1.}{$\text{Li}^+$~(1S$_0$)}} &&
    \multicolumn{4}{c}{$E_\mathrm{CC}^{(2)}$ [n$E_\mathrm{h}$]} && 
    \multicolumn{4}{c}{$E_\mathrm{CB}^{(2)}$ [$\mu E_\mathrm{h}$]} && 
    \multicolumn{4}{c}{$E_\mathrm{BB}^{(2)}$ [$\mu E_\mathrm{h}$]} \\ 
    \cline{3-6} \cline{8-11} \cline{13-16} \\[-0.35cm]
    200 && --18.7 & --18.8 & --19.2 & --19.5 && --0.227 & --0.227 & --0.228 & --0.229 &&  --99.7 &  --99.9 &  --99.9 & --100.0 \\
    300 && --19.1 & --19.1 & --19.3 & --19.5 && --0.228 & --0.228 & --0.229 & --0.229 &&  --99.9 & --100.0 & --100.0 & --100.0 \\
    400 && --19.1 & --19.1 & --19.3 & --19.5 && --0.228 & --0.228 & --0.229 & --0.229 && --101.0 & --101.0 & --101.0 & --101.1 \\
    \multicolumn{1}{r}{$\epsi_{3}$$^\text{b}$} & 
     & & & & --19.8 && & & & --0.236 && & & & --246.0 \\
    \multicolumn{1}{r}{$\epsilon_{3}$$^\text{c}$} & 
     & & & & --19.9 && & & & --0.237 && & & & \multicolumn{1}{c}{---} \\
    \hline\\[-0.35cm]
\multicolumn{1}{r}{\scalebox{1.}{$\text{Be}^{2+}$~(1S$_0$)}} &&
    \multicolumn{4}{c}{$E_\mathrm{CC}^{--}$ [n$E_\mathrm{h}$]} && 
    \multicolumn{4}{c}{$E_\mathrm{CB}^{--}$ [$\mu E_\mathrm{h}$]} && 
    \multicolumn{4}{c}{$E_\mathrm{BB}^{--}$ [$\mu E_\mathrm{h}$]} \\ 
    \cline{3-6} \cline{8-11} \cline{13-16} \\[-0.35cm]
    100 && --39.2 & --44.4 & --46.3 & --55.9 && --0.60 & --0.61 & --0.61 & --0.66 && --195 & --196 & --196 & --197 \\
    200 && --54.2 & --54.3 & --54.3 & --55.9 && --0.64 & --0.64 & --0.64 & --0.65 && --197 & --197 & --198 & --198 \\
    300 && --54.7 & --54.7 & --54.7 & --55.9 && --0.65 & --0.65 & --0.65 & --0.65 && --199 & --199 & --199 & --199 \\
    \multicolumn{1}{r}{$\epsi_{3}$$^\text{b}$} & 
     & & & & --56.7 && & & & --0.67 && & & & --188  \\
    \multicolumn{1}{r}{$\epsilon_{3}$$^\text{c}$} & 
     & & & & --56.7 && & & & --0.73 && & & & \multicolumn{1}{c}{---} \\
    \hline\hline\\[-0.35cm]
  \end{tabular}
}
  \begin{flushleft}
    $^\text{a}$ %
      $\nb$ is the number of the basis functions optimized for (the non-relativistic energy of) the reference state and $N_\mathrm{aux}$ is the auxiliary basis set optimized for $E^{(2)}_\text{CC}$, Eq.~\eqref{eq:E2spectralCC}. The listed corrections were computed by merging the reference and auxiliary basis sets. \\
    $^\text{b}$ %
      $\epsi_3$ is obtained from $\alpha$ scaling and fitting the energy correction according to Eq.~\eqref{eq:Etwoalpha}. \\
    $^\text{c}$ %
      $\epsilon_3$ is the $\alpha^3\Eh$-order nrQED value, Eqs.~\eqref{eq:epcCCmmBO} and \eqref{eq:epcCBmmBO}. 
      He (1S$_0$):  $\langle \delta(\br_{12})\rangle_\text{nr}=0.106345$; 
      He (2S$_0$): $\left\langle \delta(r_{12})\right \rangle_\mathrm{nr}=0.008648$; %
      Li$^+$: $\left\langle \delta(r_{12})\right \rangle_\mathrm{nr}=0.533723$           
      Be$^{2+}$: $\left\langle \delta(r_{12})\right \rangle_\mathrm{nr}=1.522895$ \cite{Dr06,DrYa92}. 
  \end{flushleft}    
\end{table}

\begin{table}[h]
  \caption{%
  The with-pair Dirac--Coulomb energy's $\alpha$ fine-structure constant dependence, 
  $E^\text{DC}_\text{wp}(\alpha)=\epsi_0+\epsi_2\alpha^2+\epsi_3\alpha^3+\epsi_{4,0}\alpha^4+\epsi_{4,1}\alpha^4\ln\alpha$, in $\Eh$.
  The $\epsi_k$ coefficients were obtained from fitting this function to a series of $E^\text{DC}_\text{wp}(\alpha_i)$ energies computed with $\alpha_i=(\alpha^{-1}+i)^{-1}$, $i=-50, -49, \dots, 49, 50$ and $\alpha^{-1}=137.035999177$.
  $\delta(\epsi_n)\alpha^n$ (including the $\ln\alpha$ factor for $\epsi_{4,1}$) is the difference from the nrQED value (in $\Eh$), where the nrQED value is available (Sec.~\ref{sec:nrqed}). 
    \label{tab:wpDC}
  }  
  \begin{center}
  \begin{tabular}{@{}l r r r r r r@{}} 
    \hline\hline\\[-0.35cm]
     &
    \multicolumn{1}{c}{$\epsi_0$}   &
    \multicolumn{1}{c}{$\epsi_2$} &
    \multicolumn{1}{c}{$\epsi_3$} &
    \multicolumn{1}{c}{$\epsi_{4,0}$} &
    \multicolumn{1}{c}{$\epsi_{4,1}$} \\
    \hline\\[-0.35cm] 
    Ps & 
    --0.249 999 999 999 6 \hspace{0.25cm} &  0.0468751 \hspace{0.25cm} & --0.132 649 \hspace{0.25cm} & 0.084 \hspace{0.25cm} & --0.063 5 \\
    $\delta(\epsi_n)\alpha^n$ & 
    $4\cdot 10^{-13}$ \hspace{0.25cm}  &  $5\cdot 10^{-12}$ \hspace{0.25cm} & --$8\cdot 10^{-12}$ \hspace{0.25cm} &  \multicolumn{1}{c}{---} \hspace{0.25cm} & $1\cdot 10^{-11}$ \\ \hline \\[-0.2cm]
    H & 
    --0.499 727 839 70 \hspace{0.25cm} &  --0.124 455 \hspace{0.25cm} & --0.423 80 \hspace{0.25cm} & --0.625 \hspace{0.25cm} & --1.0 \\
    $\delta(\epsi_n)\alpha^n$ & 
    $1\cdot 10^{-11}$ \hspace{0.25cm}  &  $6\cdot 10^{-11}$ \hspace{0.25cm} & $1\cdot 10^{-11}$ \hspace{0.25cm} &  \multicolumn{1}{c}{---} \hspace{0.25cm} &    \multicolumn{1}{c}{---} 
    \\ \hline \\[-0.2cm]
    Mu & 
    --0.497 593 472 911 \hspace{0.25cm} &  --0.120 227 3 \hspace{0.25cm} & --0.419 2 \hspace{0.25cm} &  --0.589 \hspace{0.25cm} & --1.0 \\
    $\delta(\epsi_n)\alpha^n$ & 
    $6\cdot 10^{-12}$ \hspace{0.25cm}  &  $8\cdot 10^{-12}$ \hspace{0.25cm} & $7\cdot 10^{-11}$ \hspace{0.25cm} &  \multicolumn{1}{c}{---} \hspace{0.25cm} &  \multicolumn{1}{c}{---} 
    \\ \hline \\[-0.2cm]
    $\mu$H & 
    --92.920 417 310 4 \hspace{0.25cm} & --8.437 64 \hspace{0.25cm} & --68.10 \hspace{0.25cm} &  --45.6 \hspace{0.25cm}  & --131 \\
    $\delta(\epsi_n)\alpha^n$ & 
    $1\cdot 10^{-9}$ \hspace{0.25cm}  &  $3\cdot 10^{-9}$ \hspace{0.25cm} & $4\cdot 10^{-9}$ \hspace{0.25cm} &  \multicolumn{1}{c}{---} \hspace{0.25cm} &  \multicolumn{1}{c}{---} 
    \\ 
    \hline \hline \\[-0.2cm]
    He & --2.903 724 376 7 \hspace{0.25cm} & --2.480 844 \hspace{0.25cm} & --0.354 52 \hspace{0.25cm} & --3.76 \hspace{0.25cm} & 0.168  \\
    $\delta(\epsi_n)\alpha^n$ & 
    $3\cdot 10^{-10}$ \hspace{0.25cm}  & $2\cdot 10^{-10}$ \hspace{0.25cm} & --$1\cdot 10^{-11}$\hspace{0.25cm} &  \multicolumn{1}{c}{---} \hspace{0.25cm} & --$2\cdot 10^{-11}$ \\ \hline \\[-0.2cm]
    He (2S$_0$) & --2.145 974 045 7 \hspace{0.25cm} & --2.079 252 \hspace{0.25cm} & --0.028 91 \hspace{0.25cm} & --4.05 \hspace{0.25cm} &   0.005 \\
    $\delta(\epsi_n)\alpha^n$ & 
    $3\cdot 10^{-10}$ \hspace{0.25cm}  & $2\cdot 10^{-10}$ \hspace{0.25cm} & --$3\cdot 10^{-11}$\hspace{0.25cm} &   \multicolumn{1}{c}{---} \hspace{0.25cm} & $1\cdot 10^{-10}$ \\ \hline \\[-0.2cm]
    Li$^+$ & --7.279 913 408 \hspace{0.25cm} & --14.734 7\hspace{0.25cm} & --1.80 \hspace{0.25cm} & --58.7 \hspace{0.25cm} &  --0.2 \\
    $\delta(\epsi_n)\alpha^n$ & 
    $6\cdot 10^{-9}$ \hspace{0.25cm}  & $5\cdot 10^{-9}$ \hspace{0.25cm} & --$7\cdot 10^{-9}$\hspace{0.25cm} &  \multicolumn{1}{c}{---} \hspace{0.25cm} & $1\cdot 10^{-8}$  \\ \hline \\[-0.2cm]
    Be$^{2+}$ & --13.655 566 235\hspace{0.25cm} & --50.485 0\hspace{0.25cm} & --5.2 \hspace{0.25cm} & --382 \hspace{0.25cm} & --5 \\
    $\delta(\epsi_n)\alpha^n$ & 
    $3\cdot 10^{-9}$ \hspace{0.25cm}  & $2\cdot 10^{-8}$ \hspace{0.25cm} & --$4\cdot 10^{-8}$\hspace{0.25cm} &    \multicolumn{1}{c}{---} \hspace{0.25cm} & $1\cdot 10^{-7}$ \\
    \hline\hline
  \end{tabular}
  \end{center}
\end{table}

\clearpage
\section*{Supplementary Material}
\noindent%
The \som\ contains:
(a) Mathematical properties of the eigenvalues and eigenfunctions of the with-pair Hamiltonian. 
(b) Calculation of $\epsilon_{3,{\rm CB}}^{--}$.
(c) Convergence tables of the non-relativistic, no-pair Dirac-Coulomb, and with-pair Dirac-Coulomb energies.
Further data are available through the Zenodo repository \cite{zenodoJeMa25}.

\vspace{0.5cm}
\begin{acknowledgments}
\noindent %
PJ gratefully acknowledges the support of the János Bolyai Research Scholarship of the Hungarian Academy of Sciences (BO/285/22). We thank the Momentum Programme of the Hungarian Academy of Sciences (LP2024-15/2024) and DKF for granting us access to the Hungarian Komondor HPC facility.
\end{acknowledgments}

%

\clearpage
\begin{center}
{\large
\textbf{Supplementary Material}
}\\[0.5cm]
{\large
\textbf{Double-pair Coulomb and Breit photon correction to the correlated relativistic energy}
} \\[0.5cm]

Péter Jeszenszki$^1$ and Edit Mátyus$^{1,\ast}$ \\
\emph{$^1$~MTA–ELTE Lendület `Momentum' Molecular Quantum electro-Dynamics Research Group,
Institute of Chemistry, Eötvös Loránd University, Pázmány Péter sétány 1/A, Budapest, H-1117, Hungary} \\
$^\ast$ edit.matyus@ttk.elte.hu
~\\[0.15cm]
(Dated: 8 December 2025)
\end{center}

\setcounter{section}{0}
\renewcommand{\thesection}{S\arabic{section}}
\setcounter{subsection}{0}
\renewcommand{\thesubsection}{S\arabic{section}.\arabic{subsection}}

\setcounter{equation}{0}
\renewcommand{\theequation}{S\arabic{equation}}

\setcounter{table}{0}
\renewcommand{\thetable}{S\arabic{table}}

\setcounter{figure}{0}
\renewcommand{\thefigure}{S\arabic{figure}}

\section{Mathematical properties of the eigenvalues and eigenfunctions of the non-hermitian with-pair Hamiltonian}
\subsection{Block structure of the with-pair Hamiltonian over the $\mcL_{++}\oplus\mcL_{--}$ and $\mcL_{+-}\oplus\mcL_{-+}$ subspaces \label{sec:wpProperties}}

The matrix representation of the with-pair eigenvalue equation can be solved using standard linear algebra packages,
\begin{align}
   H_\wp  \Phi_{\wp,i} &= E_{\wp,i} \Phi_{\wp,i} \ , \label{eq:wpeq} 
\end{align}
where 
\begin{align}
H_\wp &=  H_\np + V_\delta = h_1 + h_2 + \left( L_{++} - L_{--} \right)V_\ti  \ . \label{eq:Hwp}
\end{align}
Nevertheless, the with-pair Hamiltonian possesses several interesting properties that can be exploited to reduce the computational cost and provide deeper insight into the mathematical and physical properties of its solutions.

First of all, let us consider the block structure of the with-pair Hamiltonian, Eq.~\eqref{eq:Hwp}, over the two-particle $\mcL_{++}$ no-pair, the $\mcL_{\brra}=\mcL_{+-}\oplus \mcL_{-+}$ Brown-Ravenhall (BR) \cite{BrRa51}, and the $\mcL_{--}$ double-pair subspaces. 
For the interaction part, we have
\begin{align}
  (L_{++}-L_{--}) V_\ti
  &=
  (L_{++}-L_{--}) V_\ti (L_{++}+L_{\brra}+L_{--})
  \nonumber \\
  =
  &+L_{++} V_\ti L_{++}
  +
   L_{++} V_\ti L_{\brra}
  +
  L_{++} V_\ti L_{--}  
  \nonumber \\
  &
  -L_{--}V_\ti L_{++}
  -L_{--}V_\ti L_{\brra}
  -L_{--}V_\ti L_{--} \; .
\end{align}
$L_{\brra}$ labels the projector to the BR subspace. 

The sum of the one-particle Hamiltonians is block diagonal as $h_i$ commutes with the projection operators,
\begin{align}
  h_1+h_2
  &=
  (L_{++}+L_{\brra}+L_{--}) (h_1+h_2) (L_{++}+L_{\brra}+L_{--})
  \nonumber \\
  &=
  L_{++} (h_1+h_2) L_{++}
  +
  L_{\brra} (h_1+h_2) L_{\brra}
  +
  L_{--} (h_1+h_2) L_{--}  
\end{align}
Hence, the block structure of the with-pair Hamiltonian is
\begin{align}
  H_\wp
  =
  \left[%
  \begin{array}{@{}ccc@{}}
    L_{++}(h_1+h_2+V_\ti)L_{++} & 
    L_{++} V_\ti L_{\brra} & 
    L_{++} V_\ti L_{--} \\
    0 & 
    L_{\brra} (h_1+h_2) L_{\brra} & 
    0 \\
    -L_{--}V_\ti L_{++} &
    -L_{--}V_\ti L_{\brra} &
    L_{--} (h_1+h_2-V_\ti) L_{--} \\
  \end{array} 
  \right]  \ .
  \label{eq:blockH}
\end{align}
which can be substituted into the eigenvalue equation in Eq.~\eqref{eq:wpeq},
\begin{align}
  &%
\left[%
  \begin{array}{@{}ccc@{}}
    L_{++}(h_1+h_2+V_\ti)L_{++} & 
    L_{++} V_\ti L_{\brra} & 
    L_{++} V_\ti L_{--} \\
    0 & 
    L_{\brra} (h_1+h_2) L_{\brra} & 
    0 \\
    -L_{--}V_\ti L_{++} &
    -L_{--}V_\ti L_{\brra} &
    L_{--} (h_1+h_2-V_\ti) L_{--} \\
  \end{array} 
  \right]
  \left[%
  \begin{array}{@{}c@{}}
    \phi_{\mathrm{wp},i}^{\pupu} \\[0.10cm]
    \phi_{\mathrm{wp},i}^{\brra} \\ 
    \phi_{\mathrm{wp},i}^{\mimi} 
  \end{array}   
  \right]
  =
  E_{\mathrm{wp},i}
  \left[%
  \begin{array}{@{}c@{}}
    \phi_{\mathrm{wp},i}^{\pupu} \\[0.10cm] 
    \phi_{\mathrm{wp},i}^{\brra} \\ 
    \phi_{\mathrm{wp},i}^{\mimi} 
  \end{array}   
  \right] \; ,
  \label{eq:fullHameival} 
\end{align}
The $E_{\mathrm{wp},i}$ eigenvalues can be obtained as roots of the characteristic polynomial,
\begin{align}
   \left|%
  \begin{array}{@{}ccc@{}}
    L_{++}(h_1+h_2+V_\ti)L_{++} - E_{\mathrm{wp},i}& 
    L_{++} V_\ti L_{\brra} & 
    L_{++} V_\ti L_{--} \\
    0 & 
    L_{\brra} (h_1+h_2) L_{\brra} -  E_{\mathrm{wp},i}& 
    0 \\
    -L_{--}V_\ti L_{++} &
    -L_{--}V_\ti L_{\brra} &
    L_{--} (h_1+h_2-V_\ti) L_{--} - E_{\mathrm{wp},i}\\
  \end{array} 
  \right|  = 0 \ , 
\end{align}
and the determinant can be factorized according to its expansion about the second (block) row,
\begin{align}
  &
  \left| %
  L_{\brra} (h_1+h_2) L_{\brra} - E_{\mathrm{wp},i}  %
  \right| %
  \left|%
    \begin{array}{@{}cc@{}}
     L_{++}(h_1+h_2+V_\ti)L_{++} - E_{\mathrm{wp},i} &  L_{++} V_\ti L_{--}  \\
    -L_{--}V_\ti L_{++}   &    L_{--} (h_1+h_2-V_\ti) L_{--} - E_{\mathrm{wp},i}
    \end{array} 
  \right|=0 \ .
  \label{eq:factorization}
\end{align}
This result shows that the states from the $\mcL_{\brra}=\mcL_{+-}\oplus\mcL_{-+}$ subspace do not contribute to the energy eigenvalues of the $\mcL_{++}\oplus \mcL_{--}$ subspace. 
Hence, the effective Hamiltonian in the latter, physically relevant subspace is
\begin{align}
  H'_\wp 
  &= 
  \left(L_{++} + L_{--} \right) %
  \left( h_1 + h_2\right) %
  \left( L_{++} + L_{--}\right) 
  + \left( L_{++} - L_{--}\right) V_\ti \left( L_{++} + L_{--}\right) %
  \ . \label{eq:wpHampm}
\end{align}

At the same time, it is important to note that the left and right eigenfunctions of the non-hermitian with-pair Hamiltonian are distinct. 
Moreover, the $\mcL_{\brra}$ subspace contributes differently to the left and right eigenfunctions, although the $\mcL_{++}\oplus\mcL_{--}$ and $\mcL_{\brra}$ contributions can still be separated at the level of the eigenvalues.

\subsection{Eigenfunctions}
Regarding the right eigenfunctions of $H_\text{wp}$, we obtain the components for the $\mcL_{++}$ and $\mcL_{--}$ subspaces by solving the eigenvalue equation, 
\begin{align}
 H_\mathrm{wp}'
  \left[%
  \begin{array}{@{}c@{}}
    \phi_{\mathrm{wp},i}^{\pupu} \\ \phi_{\mathrm{wp},i}^{\mimi} 
  \end{array}   
  \right]
  =
  \left[%
    \begin{array}{@{}cc@{}}
     L_{++}(h_1+h_2+V_\ti)L_{++} &  L_{++} V_\ti L_{--}  \\
    -L_{--}V_\ti L_{++}   &    L_{--} (h_1+h_2-V_\ti) L_{--} 
    \end{array} 
  \right]
  \left[%
  \begin{array}{@{}c@{}}
    \phi_{\mathrm{wp},i}^{\pupu} \\ \phi_{\mathrm{wp},i}^{\mimi} 
  \end{array}   
  \right]
  =
   E_{\mathrm{wp},i}
  \left[%
  \begin{array}{@{}c@{}}
    \phi_{\mathrm{wp},i}^{\pupu} \\ \phi_{\mathrm{wp},i}^{\mimi} 
  \end{array}   
  \right] 
%
 \;. \label{eq:Hwppeig}
\end{align}
Then, by substituting $(E_{\mathrm{wp},i}, \phi_{\mathrm{wp},i}^{\pupu}, \phi_{\mathrm{wp},i}^{\mimi})$ 
into the eigenvalue equation of the full Hamiltonian, Eq.~\eqref{eq:fullHameival}, 
we obtain
\begin{align} 
  \phi_{\mathrm{wp},i}^{\brra}=0 \; .
  \label{eq:rightbrra}
\end{align}

Regarding the left eigenfunctions of $H_\text{wp}$, labeled by $\lphi_i$, we consider
\begin{align}
  \left[%
  \begin{array}{@{}ccc@{}}
    \lphi_{\mathrm{wp},i}^{\pupu} ,
    \lphi_{\mathrm{wp},i}^{\brra} ,
    \lphi_{\mathrm{wp},i}^{\mimi} 
  \end{array}   
  \right] H_\mathrm{wp}
  =
  \left[%
  \begin{array}{@{}ccc@{}}
    \lphi_i^{\pupu} ,
    \lphi_i^{\brra} ,
    \lphi_i^{\mimi} 
  \end{array}   
  \right]  
   E_{\mathrm{wp},i} \; .
\end{align}
First, let us solve, 
\begin{align}
  \left[%
  \begin{array}{@{}cc@{}}
    \lphi_{\mathrm{wp},i}^{\pupu} ,
    \lphi_{\mathrm{wp},i}^{\mimi} 
  \end{array}   
  \right]
  \left[%
    \begin{array}{@{}cc@{}}
     L_{++}(h_1+h_2+V_\ti)L_{++} &  L_{++} V_\ti L_{--}  \\
    -L_{--}V_\ti L_{++}   &    L_{--} (h_1+h_2-V_\ti) L_{--} 
    \end{array} 
  \right]
  =
  \left[%
  \begin{array}{@{}ccc@{}}
    \lphi_{\mathrm{wp},i}^{\pupu} ,
    \lphi_{\mathrm{wp},i}^{\mimi} 
  \end{array}   
  \right]  
   E_{\mathrm{wp},i} 
  \;, 
\end{align}
and then, substituting these results into the BR block in Eq. \eqref{eq:fullHameival},
\begin{align}
  \lphi_{\mathrm{wp},i}^{\brra}  E_{\mathrm{wp},i}
  &=
  \lphi_{\mathrm{wp},i}^{\pupu} V_i L_\brra 
  + 
  \lphi_{\mathrm{wp},i}^{\brra} \left( h_1+h_2\right)\mcL_\brra 
  -
  \lphi_{\mathrm{wp},i}^{\mimi} V_i L_\brra \\
  \lphi_{\mathrm{wp},i}^{\brra}\left[ E_{\mathrm{wp},i} - \left(h_1+h_2\right)L_\brra \right]
  &=
  \lphi_{\mathrm{wp},i}^{\pupu} V_i L_\brra
  -
  \lphi_{\mathrm{wp},i}^{\mimi}V_i L_\brra 
  \nonumber \\
  \Rightarrow\quad %
  \lphi_{\mathrm{wp},i}^{\brra} 
  &=
  [\lphi_{\mathrm{wp},i}^{\pupu}  
  -
  \lphi_{\mathrm{wp},i}^{\mimi} ] V_i L_\brra ( E_{\mathrm{wp},i} - h_1-h_2)^{-1} L_\brra \; .
  \label{eq:leftbrra}
\end{align}
The subindex $i$ indicates that the left eigenfunction $\lphi_{\mathrm{wp},i}$ and the right eigenfunction $\phi_{\mathrm{wp},i}$ correspond to the eigenvalue $E_{\mathrm{wp},i}$, which is determined by solely the $\mcL_{++}\oplus \mcL_{--}$ subspace. 
We emphasise, however, that the eigenfunctions are defined over the entire Hilbert space, $\mcL_{++}\oplus \mcL_{\brra}\oplus \mcL_{--}$. 
This full dependence---including the $\mcL_{\brra}$ subspace, which does not affect $E_{\mathrm{wp},i}$---becomes relevant in the computation of energy corrections beyond the instantaneous photon exchange.

\subsection{Eigenvalues}
Based on the factorization in Eq.~\eqref{eq:factorization}, the eigenvalues associated with the $\mcL_{\brra}$ subspace (denoted by the $\tildi$ index) can be obtained by solving
\begin{align}
  {H}^{\brra,\brra}_\mathrm{wp} {\phi}_{\mathrm{wp},\tildi}^{\brra} 
  = 
   E_{\mathrm{wp},\tildi}
  {\phi}_{\mathrm{wp},\tildi}^{\brra}
  \quad 
  \Rightarrow
  \quad 
  (E_{\mathrm{wp},\tildi}, {\phi}_{\mathrm{wp},\tildi}^{\brra}) \; .
  \label{eq:brE}  
\end{align}
Due to the hermiticity of the $H^{\brra,\brra}_\mathrm{wp}$ block, 
\begin{align}
  H^{\brra,\brra}_\mathrm{wp} = L_{\brra} (h_1+h_2) L_{\brra} \; ,
\end{align}
the $E_{\mathrm{wp},\tildi}$ eigenvalues are real and
\begin{align}
  \lphi_{\mathrm{wp},\tildi}^{\brra} = ({\phi}_{\mathrm{wp},\tildi}^{\brra})^\dagger \; .
\end{align}
Then, the (right and left) eigenfunctions in the $\mcL_{++}\oplus \mcL_{--}$ subspace can be obtained by substituting $(E_{\mathrm{wp},\tildi}, \phi_{\mathrm{wp},\tildi}^{\brra})$, calculated from Eq.~\eqref{eq:brE}, into Eq.~\eqref{eq:fullHameival}, which yields a system of linear equations for $\phi^{\pupu}_{\mathrm{wp},\tildi}$ and $\phi^{\mimi}_{\mathrm{wp},\tildi}$,
\begin{align}
  (H^{\pupu,\pupu}_\mathrm{wp} - E_{\mathrm{wp},\tildi}) \phi^{\pupu}_{\mathrm{wp},\tildi}
  + 
  H^{\pupu,\mimi}_\mathrm{wp} \phi^{\mimi}_{\mathrm{wp},\tildi}
  &=
  -H^{\pupu,\brra}_\mathrm{wp} \phi^{\brra}_{\mathrm{wp},\tildi}
  \nonumber \\
  -(H_\mathrm{wp}^{\pupu,\mimi})^\dagger \phi^{\pupu}_{\mathrm{wp},\tildi}
  +
  (H_\mathrm{wp}^{\mimi,\mimi}-E_{\mathrm{wp},\tildi}) \phi^{\mimi}_{\mathrm{wp},\tildi}
  &=
  -H_\mathrm{wp}^{\mimi,\brra} \phi^{\brra}_{\mathrm{wp},\tildi}
  \quad\Rightarrow\quad %
  (\phi^{\pupu}_{\mathrm{wp},\tildi},\phi^{\mimi}_{\mathrm{wp},\tildi}) \; .
\end{align}
The left eigenfunctions can be obtained by substituting $(E_{\mathrm{wp},\tildi},\lphi_{\mathrm{wp},\tildi}^{\brra})$ into the left eigenvalue equation of the full Hamiltonian, which results in
\begin{align}
  \lphi^{\pupu}_{\mathrm{wp},\tildi} = 0 
  \quad \text{and} \quad
  \lphi^{\mimi}_{\mathrm{wp},\tildi} = 0  \; .
\end{align}

In summary, the with-pair Hamiltonian corresponding to instantaneous interactions has the following eigenvalues and eigenfunctions 
\begin{align}
  E_{\mathrm{wp},i} \; ,
  \quad 
  \lphi_{\mathrm{wp},i} = 
  \left[%
  \begin{array}{@{}c@{}}
    \lphi^{\pupu}_{\mathrm{wp},i} ,
    \lphi^{\brra}_{\mathrm{wp},i} ,
    \lphi^{\mimi}_{\mathrm{wp},i}
  \end{array}
  \right]  \; ,
  \quad
  \phi_{\mathrm{wp},i} = 
  \left[%
  \begin{array}{@{}c@{}}
    \phi^{\pupu}_{\mathrm{wp},i}
    \\
    0 
    \\
    \phi^{\mimi}_{\mathrm{wp},i}
  \end{array}
  \right]
  \; , 
  \quad i \in \mathcal{I}_{{\pupu},{\mimi}} \; ,
  \label{eq:eigenpm}
\end{align}
and
\begin{align}
  E_{\mathrm{wp}, \tildi} \; ,
  \quad 
  \lphi_{\mathrm{wp},\tildi} = 
  \left[%
  \begin{array}{@{}c@{}}
    0 ,
    \lphi^{\brra}_{\mathrm{wp},\tildi} ,
    0
  \end{array}
  \right] \; ,
  \quad
  \phi_{\mathrm{wp},\tildi} = 
  \left[%
  \begin{array}{@{}c@{}}
    \phi^{\pupu}_{\mathrm{wp},\tildi} \\[0.10cm]
    \phi^{\brra}_{\mathrm{wp},\tildi} \\
    \phi^{\mimi}_{\mathrm{wp},\tildi}
  \end{array}
  \right]
  \; , \quad \tildi \in \mathcal{I}_{\brra} \; .
  \label{eq:eigenbr}
\end{align}
Different index symbols are used to distinguish the various subspaces; $i$ and $\tildi$ may be either discrete or continuous, with the corresponding eigenvalues determined by the 
($\mcL_{++}\oplus\mcL_{--}$ and $\mcL_{+-}\oplus\mcL_{-+}$) subspaces, respectively.

\subsection{Relation of the left and right eigenfunctions of the with-pair Hamiltonian, particle and charge expectation value \label{sec:leftright}}
We start with the $\mcL_{\pupu} \oplus \mcL_{\mimi}$ subspace, and consider the correspondig $H'_\wp$ Hamiltonian introduced in Eq.~\eqref{eq:wpHampm}. 
We note the anti-hermitian relation of the off-diagonal blocks,
$H^{--,++}_\mathrm{wp} = (H^{++,--}_\mathrm{wp})^\dagger$, and study its effect on the left and right eigenfunctions.
Let us assume that we have found a right eigenfunction, which fulfills the following system of equations,
\begin{align}
  H_\mathrm{wp}^{\pupu,\pupu} \, \phi_i^{\mathrm{wp},\pupu} 
  &+ 
  H_\mathrm{wp}^{\pupu,\mimi} \, \phi_{\mathrm{wp},i}^{\mimi} 
  =
  E_{\mathrm{wp},i} \, \phi_{\mathrm{wp},i}^{\pupu}  \label{eq:rightfun1} \\
  -(H_\mathrm{wp}^{\pupu,\mimi})^\dagger \, \phi_{\mathrm{wp},i}^{\pupu} 
  &+
  H_\mathrm{wp}^{\mimi,\mimi} \, \phi_{\mathrm{wp},i}^{\mimi} 
  =
  E_{\mathrm{wp},i} \, \phi_{\mathrm{wp},i}^{\mimi} \; ,
  \label{eq:rightfun2}
\end{align}
and the left eigenfunction with the same $E_{\mathrm{wp},i}$ energy, which fulfills
\begin{align}
  \lphi^{\pupu}_{\mathrm{wp},i} \, H_\mathrm{wp}^{\pupu,\pupu} 
  &- 
  \lphi^{\mimi}_{\mathrm{wp},i} \, {H_\mathrm{wp}^{\pupu,\mimi}}^\dagger
  =
  E_{\mathrm{wp},i} \, \lphi^{\pupu}_{\mathrm{wp},i} \\
  \lphi^{\pupu}_{\mathrm{wp},i} \, H_\mathrm{wp}^{\pupu,\mimi} 
  &+ 
  \lphi^{\mimi}_{\mathrm{wp},i} \, H_\mathrm{wp}^{\mimi,\mimi} \, \,
  =
  E_{\mathrm{wp},i} \, \lphi^{\mimi}_{\mathrm{wp},i} \; .
  \label{eq:leftfun}
\end{align}
Then, by exploiting the hermiticity of the diagonal blocks, 
${H_\mathrm{wp}^{\pupu,\pupu}}^\dagger=H_\mathrm{wp}^{\pupu,\pupu}$ and ${H_\mathrm{wp}^{\mimi,\mimi}}^\dagger=H_\mathrm{wp}^{\mimi,\mimi}$,
we obtain
\begin{align}
  H_\mathrm{wp}^{\pupu,\pupu} \, {\lphi^{\pupu}_{\mathrm{wp},i}}{}^\dagger
  &- 
  {H_\mathrm{wp}^{\pupu,\mimi}} \, {\lphi^{\mimi}_{\mathrm{wp},i}}{}^\dagger 
  =
  E_{\mathrm{wp},i} \, {\lphi^{\pupu}_{\mathrm{wp},i}}{}^\dagger \\
  H^{\pupu,\mimi}_\mathrm{wp}{}^\dagger \, {\lphi^{\pupu}_{\mathrm{wp},i}}{}^\dagger
  &+ 
  H^{\mimi,\mimi}_\mathrm{wp} \, {\lphi^{\mimi}_{\mathrm{wp},i}}{}^\dagger \, \, 
  =
  E_{\mathrm{wp},i} \, {\lphi^{\mimi}_{\mathrm{wp},i}}{}^\dagger \ ,
  \label{eq:leftfundagger}
\end{align}
which in comparison with Eqs.~\eqref{eq:rightfun1} and \eqref{eq:rightfun2} provide the following relations for the left and right eigenfunctions:
\begin{align}
  {\lphi^\pupu_{\mathrm{wp},i}}{}^\dagger = \phi^\pupu_{\mathrm{wp},i} 
  \, \, \, \, \, \, \quad\text{and}\quad
  {\lphi^\mimi_{\mathrm{wp},i}}{}^\dagger = -\phi^\mimi_{\mathrm{wp},i} 
  \label{eq:sign1}
\end{align}
or
\begin{align}
  {\lphi^\pupu_{\mathrm{wp},i}}{}^\dagger = -\phi^\pupu_{\mathrm{wp},i} 
   \quad\text{and}\quad
  {\lphi^\mimi_{\mathrm{wp},i}}{}^\dagger = \phi^\mimi_{\mathrm{wp},i}  \; .
  \label{eq:sign2} 
\end{align}
Out of the two options, the normalization condition 
\begin{align}
  \lphi^\pupu_{\mathrm{wp},i} \, \phi^\pupu_{\mathrm{wp},i}
  +
  \lphi^\mimi_{\mathrm{wp},i} \, \phi^\mimi_{\mathrm{wp},i} 
  =
  1
\end{align}
fixes the actual sign relations, Eqs.~\eqref{eq:sign1} or \eqref{eq:sign2}.
These results can be summarized, by adopting a bra-ket notation, 
\begin{align}
  1
  =
  \brapsket{\phi^\pupu_{\mathrm{wp},i}}{-\phi^\mimi_{\mathrm{wp},i}}{\phi^\pupu_{\mathrm{wp},i}}{\phi^\mimi_{\mathrm{wp},i}} 
  =
  \langle \phi^\pupu_{\mathrm{wp},i} | \phi^\pupu_{\mathrm{wp},i} \rangle
  -
  \langle \phi^\mimi_{\mathrm{wp},i} | \phi^\mimi_{\mathrm{wp},i} \rangle \; ,
  \quad
  \text{if}\ %
  \langle \phi^\pupu_{\mathrm{wp},i} | \phi^\pupu_{\mathrm{wp},i} \rangle 
  > 
  \langle \phi^\mimi_{\mathrm{wp},i} | \phi^\mimi_{\mathrm{wp},i} \rangle \; ,
\end{align}
\emph{i.e.,} for `electronic' solutions,
whereas 
\begin{align}
  1
  =
  \brapsket{-\phi^\pupu_{\mathrm{wp},i}}{\phi^\mimi_{\mathrm{wp},i}}{\phi^\pupu_{\mathrm{wp},i}}{\phi^\mimi_{\mathrm{wp},i}} 
  =
  -\langle \phi^\pupu_{\mathrm{wp},i} | \phi^\pupu_{\mathrm{wp},i} \rangle
  +
  \langle \phi^\mimi_{\mathrm{wp},i} | \phi^\mimi_{\mathrm{wp},i} \rangle \; ,
  \quad
  \text{if}\ 
  \langle \phi^\mimi_{\mathrm{wp},i} | \phi^\mimi_{\mathrm{wp},i} \rangle 
  > 
  \langle \phi^\pupu_{\mathrm{wp},i} | \phi^\pupu_{\mathrm{wp},i} \rangle \; ,
\end{align}
\emph{i.e.,} for `positronic' solutions.

In addition to the eigenfunction blocks corresponding to the subspace $\mcL_{\pupu}\oplus\mcL_{\mimi}$,
the BR block is zero in the right eigenfunction, Eq.~\eqref{eq:rightbrra} and calculable from Eq.~\eqref{eq:leftbrra} for the left eigenfunction. 
Thus, the eigenfunctions normalized to 1 over the entire Hilbert space can be written as
\begin{align}
  \braharom{\phi^\pupu_{\mathrm{wp},i}}{\phi^\brra_{\mathrm{wp},i}}{-\phi^\mimi_{\mathrm{wp},i}}
  \quad
  \ketharom{\phi^\pupu_{\mathrm{wp},i}}{0}{\phi^\mimi_{\mathrm{wp},i}}  
  \quad \text{for `electronic':}\quad
  |\phi^\pupu_{\mathrm{wp},i}|^2
  > 
  |\phi^\mimi_{\mathrm{wp},i}|^2 \label{eq:eivecelect}
\end{align}
and 
\begin{align}
  \braharom{-\phi^\pupu_{\mathrm{wp},i}}{\phi^\brra_{\mathrm{wp},i}}{\phi^\mimi_{\mathrm{wp},i}}
  \quad
  \ketharom{\phi^\pupu_{\mathrm{wp},i}}{0}{\phi^\mimi_{\mathrm{wp},i}}  
  \quad \text{for `positronic':}\quad
  |\phi^\mimi_{\mathrm{wp},i}|^2
  >
  |\phi^\pupu_{\mathrm{wp},i}|^2 \label{eq:eivec2b}
\end{align}
We see that the BR block contributes only to the left eigenfunctions, but its contribution to the right eigenfunctions is zero. 

For electronic (positronic) solutions, there is a nonvanishing contribution also from the $\mcL_{\mimi}$ ($\mcL_{\pupu}$) subspace, $|\phi^\mimi_{\mathrm{wp},i}|^2>0$ ($|\phi^\pupu_{\mathrm{wp},i}|^2$), and thus the normalization condition can be fulfilled only if $|\phi^\pupu_{\mathrm{wp},i}|^2>1$ ($|\phi^\mimi_{\mathrm{wp},i}|^2>1$). The positronic contribution to full with-pair wave functions norm is relatively small for the systems studied, numerical values (for the largest basis results) are shownin Table~\ref{tab:eta}.

\begin{table}[]
  \centering
  \caption{%
    Ratio of the negative- and positive-energy space contributions (norm over these subspaces) of the with-pair eigenfunction, $\eta=\left| \phi_\wp^{--} \right|/\left| \phi_\wp^{++} \right|$, calculated with the largest basis sets. The values are converged to the digits printed.
    \label{tab:eta}    
  }
  \begin{tabular}{@{}l r r r r@{}} 
    \hline\hline\\[-0.35cm]
     &
    \multicolumn{1}{c}{$\text{Ps}$}   &
    \multicolumn{1}{c}{H} &
    \multicolumn{1}{c}{Mu} &
    \multicolumn{1}{c}{$\mu$H}   \\
    \hline\\[-0.35cm] 
    $\eta$ & 
    $1.1 \cdot 10^{-7}$ &   
    $1.8 \cdot 10^{-10}$ \hspace{0.25cm} & 
    $2.8 \cdot 10^{-12}$ \hspace{0.25cm} & 
    $2.6 \cdot 10^{-8}$ \hspace{0.25cm}  \\
    \hline \\
     & 
    \multicolumn{1}{c}{He (1S$_0$)} &
    \multicolumn{1}{c}{He (2S$_0$)} &
    \multicolumn{1}{c}{Li$^+$ (1S$_0$)} &
    \multicolumn{1}{c}{Be$^{2+}$ (1S$_0$)} \\ \hline \\[-0.35cm] 
    $\eta$ & 
    $1.8 \cdot 10^{-7}$ \hspace{0.25cm} & 
    $5.0 \cdot 10^{-8}$ \hspace{0.25cm} & 
    $4.0 \cdot 10^{-7}$ \hspace{0.25cm} & 
    $6.8 \cdot 10^{-7}$ \\
    \hline\hline
  \end{tabular}
\end{table}

\subsection{Realness of the eigenvalues}

Araki \cite{Ar57} has already shown that for Coulomb interactions, the eigenvalues of the Hamiltonian with pairs are real. In this subsection, his derivation is presented in greater detail. From the calculation, it can be seen that it remains valid for any instantaneous interactions and also for pBO systems.

The energy can be calculated by projecting the eigenvalue equation, Eq.~\eqref{eq:Hwppeig}, from the left with the left eigenfunction in Eqs.~\eqref{eq:eivecelect} and \eqref{eq:eivec2b},
\begin{align}
E_{\mathrm{wp},i}' = 
\left[  \phi_{\mathrm{wp},i}^{\pupu}, - \phi_{\mathrm{wp},i}^{\mimi} \right]    H_\mathrm{wp}'
  \left[%
  \begin{array}{@{}c@{}}
    \phi_{\mathrm{wp},i}^{\pupu} \\ \phi_{\mathrm{wp},i}^{\mimi}
  \end{array}   
  \right]  \ , \label{eq:expval}
\end{align}
where $E_{\mathrm{wp},i}'=E_{\mathrm{wp},i}$ for the `electronic' solution and $E_{\mathrm{wp},i}'=-E_{\mathrm{wp},i}$ for the `positronic' solution (compare Eqs.~\eqref{eq:eivecelect} and \eqref{eq:eivec2b}).
Using the relation $\left[  \phi_{\mathrm{wp},i}^{\pupu}, - \phi_{\mathrm{wp},i}^{\mimi} \right]=\left[  \phi_{\mathrm{wp},i}^{\pupu}, \phi_{\mathrm{wp},i}^{\mimi} \right]({L}_{++}-{L}_{--})$ and Eq.~\eqref{eq:wpHampm}, the energy in Eq.~\eqref{eq:expval} can be written as
\begin{align}
    E_{\mathrm{wp},i}' =& \left[  \phi_{\mathrm{wp},i}^{\pupu},  \phi_{\mathrm{wp},i}^{\mimi} \right] \bigg[ \left(L_{++} - L_{--} \right) %
  \left( h_1 + h_2\right) %
  \left( L_{++} + L_{--}\right) 
  + \nonumber \\ 
  & \hspace{5cm} \left( L_{++} + L_{--}\right) V_\ti \left( L_{++} + L_{--}\right) \bigg] \left[%
  \begin{array}{@{}c@{}}
    \phi_{\mathrm{wp},i}^{\pupu} \\ \phi_{\mathrm{wp},i}^{\mimi}
  \end{array}   
  \right] \ .
\end{align}
As $[h_i,L_{++}]=[h_i,L_{--}]=0$ and $L_{++}L_{--}=L_{--}L_{++}=0$, we obtain
\begin{align}
    E_{\mathrm{wp},i}' =& \left[  \phi_{\mathrm{wp},i}^{\pupu},  \phi_{\mathrm{wp},i}^{\mimi} \right] \bigg[ L_{++}  %
  \left( h_1 + h_2\right)  L_{++} 
  - L_{--}\left( h_1 + h_2\right) L_{--}  
  + \nonumber \\ 
  & \hspace{5cm} \left( L_{++} + L_{--}\right) V_\ti \left( L_{++} + L_{--}\right) \bigg] \left[%
  \begin{array}{@{}c@{}}
    \phi_{\mathrm{wp},i}^{\pupu} \\ \phi_{\mathrm{wp},i}^{\mimi}
  \end{array}   
  \right] \ , \\
  E_{\mathrm{wp},i}' =& \left \langle \phi_{\mathrm{wp},i}^{\pupu} \left| h_1+ h_2 \right|  \phi_{\mathrm{wp},i}^{\pupu}  \right \rangle - \left \langle \phi_{\mathrm{wp},i}^{\mimi} \left| h_1+ h_2 \right| \phi_{\mathrm{wp},i}^{\mimi}  \right \rangle  
  + \left[  \phi_{\mathrm{wp},i}^{\pupu},  \phi_{\mathrm{wp},i}^{\mimi} \right] V_i 
   \left[%
  \begin{array}{@{}c@{}}
    \phi_{\mathrm{wp},i}^{\pupu} \\ \phi_{\mathrm{wp},i}^{\mimi}
  \end{array}   
  \right] \ ,
\end{align}
which is the expectation value of a hermitian operator. 
Hence, $E_{\mathrm{wp},i}'$ ($E_{\mathrm{wp},i}$) is real.

\clearpage
\section{Calculation of $\epsilon_{3,{\rm CB}}^{--}$}
In many aspects, the derivation follows the calculation of $\epsilon_{3,\text{CC}}^{--}$ in Ref.~\citenum{sucherPhD1958}. Our calculation can be used for two-electron BO systems or for genuine two spin-1/2-fermion systems with equal masses, $m=m_1=m_2$.

We start by considering $E_\mathrm{CB}^{(2)}$ in 
Eq.~(30) of the main text,  
\begin{align}
  E_\mathrm{CB}^{(2)}
  &= 
  \text{Re}
  \left[ 
    \sum_{m=1}^{N}
    \frac{%
      \langle \phi^{++}_{\np}|V_\mathrm{C}|\phi^{--}_{0,m} \rangle
      \langle \phi^{--}_{0,m}|V_\mathrm{B} |\phi^{++}_{\np} \rangle 
    }{E^{--}_{0,m}-E^{++}_{\np} } 
  \right] \, , \label{eq:ECBorig}
\end{align}
and rewrite the spectral representation of the resolvent to operator form
\begin{align}
   \sum_{m=1}^{N}
   \frac{%
     |\phi^{--}_{0,m} \rangle \langle \phi^{--}_{0,m}|
   }{%
     E^{--}_{0,m}-E^{++}_{\np}
   } 
   =
   -\frac{L_{--}}{E^{++}_\mathrm{\np}-h_1-h_2} \ ,
\end{align}

Then, we use the momentum-space representation of the Coulomb, $V_\text{C}$, and the Breit, $V_\text{B}$, interactions, Eqs.~(5) and (6), and then, $E_\mathrm{CB}^{(2)}$ in Eq.~\eqref{eq:ECBorig} can be written as 
\begin{align}
   E_\mathrm{CB}^{(2)}
   &=
   -\frac{z_1^2 z_2^2}{4\pi^4} \, 
   \text{Re} %
   \bigg[%
   \int %
     \text{d}\bos{p}_1 \text{d}\bos{p}_2 
     \text{d}\bos{k} \text{d} \bos{k}'\ 
       [\phi_\mathrm{\np}^{++}(\bos{p}_1,\bos{p}_2)]^\dagger
       \left(-\frac{\tilde{\bos{\alpha}}_1 \tilde{\bos{\alpha}}_2}{k^2} \right) \cdot  
   \nonumber \\ 
   & \hspace{1cm} 
   \cdot 
     \frac{%
       L_{--}(\bos{p}_1-\bos{k},\bos{p}_2+\bos{k})
     }{%
       E^{++}_{\np} -h_1(\bos{p}_1-\bos{k})-h_2(\bos{p}_2+\bos{k})
     } 
     \frac{1}{k'^2} 
     \phi^{++}_\mathrm{\np}(\bos{p}_1-\bos{k}-\bos{k}',\bos{p}_2+\bos{k}+\bos{k}') 
   \bigg] \ ,
  \label{eq:ebcmmdef} 
\end{align}
where $\tilde{\bos{\alpha}}_i\tilde{\bos{\alpha}}_j = \sum_{a,b=1}^3 (\delta_{ab}-\frac{k_a k_b}{k^2}) \alpha_{ia} \alpha_{jb}$. 

The derivation mainly follows Sucher’s original work on the Coulomb-transverse term in Chapter~5 of Ref.~\citenum{sucherPhD1958}. 
For easier comparison with the expressions, we continue in natural units, and at the end of this section we convert everything back to atomic units. 
Then,
\begin{align}
   E_\mathrm{CB}^{(2)}
   &=
   -\frac{z_1^2 z_2^2e^4}{4\pi^4} \, 
   \text{Re} %
   \bigg[%
   \int %
     \text{d}\bos{p}_1 \text{d}\bos{p}_2 
     \text{d}\bos{k} \text{d} \bos{k}'\ 
       [\phi_\mathrm{\np}^{++}(\bos{p}_1,\bos{p}_2)]^\dagger
       \left(-\frac{\tilde{\bos{\alpha}}_1 \tilde{\bos{\alpha}}_2}{k^2} \right) \cdot  
   \nonumber \\ 
   & \hspace{1cm} 
   \cdot 
     \frac{%
       L_{--}(\bos{p}_1-\bos{k},\bos{p}_2+\bos{k})
     }{%
       E^{++}_{\np} -h_1(\bos{p}_1-\bos{k})-h_2(\bos{p}_2+\bos{k})
     } 
     \frac{1}{k'^2} 
     \phi^{++}_\mathrm{\np}(\bos{p}_1-\bos{k}-\bos{k}',\bos{p}_2+\bos{k}+\bos{k}') 
   \bigg] \ ,
  \label{eq:ebcmmdefn} 
\end{align} 
and $e$ is the elementary charge.

To obtain the leading-order contribution to $E_\mathrm{CB}^{(2)}$,
we approximate $L_{--}$ with the free-particle projectors, $L_{--} \approx \Lambda_{1-} \Lambda_{2-}$,
\begin{align}
  \epsilon_{\mathrm{CB}}^{--}
  &=  
  \frac{z_1^2 z_2^2 e^4}{4\pi^4} \, 
  \text{Re} %
  \bigg[%
  \int %
    \text{d}\bos{p}_1 \text{d}\bos{p}_2 
    \text{d}\bos{k} \text{d} \bos{k}'\ 
      [\phi^{++}_\np(\bos{p}_1,\bos{p}_2)]^\dagger 
      \frac{%
        \tilde{\bos{\alpha}}_1 \tilde{\bos{\alpha}}_2
      }{k^2}\ \cdot 
      \nonumber \\
    & \hspace{1cm}
    \cdot  
    \frac{%
      \Lambda_{1-}(\bos{p}_1-\bos{k}) \Lambda_{2-}(\bos{p}_2+\bos{k})
    }{%
      E_\np^{++} + E_1^0(\bos{p}_1-\bos{k}) + E_2^0(\bos{p}_2+\bos{k})
    } 
    \frac{1}{k'^2} 
    \phi^{++}_\np(\bos{p}_1-\bos{k}-\bos{k}',\bos{p}_2+\bos{k}+\bos{k}') 
  \bigg] \; ,
\end{align}
where 
\begin{align}
  h_i(\bos{p}_i) \Lambda_{i-} = -E_i^0(\bos{p}_i) \Lambda_{i-} = -\sqrt{\bos{p}_i^2+m^2}\ \Lambda_{i-} \ .
\end{align}
Next, $\phi^{++}_\np$ is approximated with the Pauli wave function, $\phi_{\rm P}$, and $E_\np^{++}$ with $2m$,
\begin{align}
  \epsilon_{\mathrm{CB}}^{--}
  &= 
  \frac{z_1^2 z_2^2 e^4}{4\pi^4} \, 
  \text{Re} 
  \bigg[%
  \int 
    \text{d}\bos{p}_1 \text{d}\bos{p}_2 
    \text{d}\bos{k} \text{d} \bos{k}'\ 
      \phi_{\rm P}^\dagger(\bos{p}_1,\bos{p}_2) 
      \frac{%
        \tilde{\bos{\alpha}}_1 \tilde{\bos{\alpha}}_2
       }{k^2}\ \cdot
      \nonumber \\ 
    & \hspace{1cm}  
    \cdot
    \frac{%
      \Lambda_{1-}(\bos{p}_1-\bos{k}) \Lambda_{2-}(\bos{p}_2+\bos{k})
    }{%
      2m + E_1^0(\bos{p}_1-\bos{k}) + E_2^0(\bos{p}_2+\bos{k})
    }       
    \frac{1}{k'^2} 
    \phi_{\rm P}(\bos{p}_1-\bos{k}-\bos{k}',\bos{p}_2+\bos{k}+\bos{k}') 
  \bigg] 
  \ .
\end{align}
The relevant part of the integral comes from the $\left|\bos{k}\right|\gg m $ region, hence, $E_1^0(\bos{p}_1-\bos{k}) \approx E_2^0(\bos{p}_2+\bos{k}) \approx E^0(\bos{k})=\sqrt{k^2+m^2}$,
\begin{align}
  \epsilon_{\mathrm{CB}}^{--}
  &= 
  \frac{z_1^2 z_2^2 e^4}{4\pi^4} \, %
  \text{Re} %
  \bigg[ %
  \int %
    \text{d}\bos{p}_1 \text{d}\bos{p}_2 
    \text{d}\bos{k} \text{d} \bos{k}'\   
      \phi_{\rm P}^\dagger(\bos{p}_1,\bos{p}_2) 
      \frac{%
        \tilde{\bos{\alpha}}_1 \tilde{\bos{\alpha}}_2
      }{k^2} \cdot 
      \nonumber \\
      & \hspace{1cm} \cdot
      \frac{%
        \Lambda_{1-}(-\bos{k}) \Lambda_{2-}(\bos{k})
        }{%
          2m+2E^0(\bos{k})
        }        
      \frac{1}{k'^2} 
      \phi_{\rm P}(\bos{p}_1-\bos{k}-\bos{k}',\bos{p}_2+\bos{k}+\bos{k}')  
    \bigg] 
    \ . 
\end{align}
Following Ref.~\citenum{sucherPhD1958}, we use the high-momentum approximation, $\bos{k} \approx -\bos{k}'$ for the potentials, 
\begin{align}
  \epsilon_{\mathrm{CB}}^{--}
  &= 
  \frac{z_1^2 z_2^2 e^4}{4\pi^4} \,  
  \text{Re} 
  \bigg[%
  \int %
    \text{d}\bos{p}_1 \text{d}\bos{p}_2 \text{d}\bos{k} \   
      \phi_{\rm P}^\dagger(\bos{p}_1,\bos{p}_2) 
      \frac{\tilde{\bos{\alpha}}_1 \tilde{\bos{\alpha}}_2}{k^4} 
      \frac{\Lambda_{1-}(-\bos{k}) \Lambda_{2-}(\bos{k})}{2m + 2E^0(\bos{k})}  \cdot \\
     & \hspace{5cm}   
     \int \text{d} \bos{k}' 
     \phi_{\rm P}(\bos{p}_1-\bos{k}-\bos{k}',\bos{p}_2+\bos{k}+\bos{k}') 
  \bigg]  \, 
  \ . \nonumber
\end{align}
Transforming the Pauli wave function to coordinate space,
\begin{align}
  \phi_\text{P}(\bos{p}_1,\bos{p}_2)
  = 
  \frac{1}{(2 \pi)^3} 
  \int \mbox{d} \bos{r}_1 \mbox{d} \bos{r}_2\ 
    \eem^{\iim \left(\bos{p}_1\bos{r}_1+\bos{p}_2 \bos{r}_2\right)}  
    \phi_{\rm P}(\bos{r}_1,\bos{r}_2) \  , 
\end{align}
and integrating over $\bos{p}_1$ and  $\bos{p}_2$,
\begin{align}
  \epsilon_{\mathrm{CB}}^{--} 
  &= 
  \frac{z_1^2 z_2^2 e^4}{4\pi^4}\,  
  \text{Re} 
  \bigg[%
  \int %
    \mbox{d} \bos{r}_1 \mbox{d} \bos{r}_2 \mbox{d} \bos{k}\  
    \phi_{\rm P}^\dagger(\bos{r}_1,\bos{r}_2) 
    \frac{\tilde{\bos{\alpha}}_1 \tilde{\bos{\alpha}}_2}{k^4} 
    \frac{\Lambda_{1-}(-\bos{k}) \Lambda_{2-}(\bos{k})}{2m + 2E^0(\bos{k})}  
    \phi_{\rm P}(\bos{r}_1,\bos{r}_2) 
  \bigg] \cdot 
  \nonumber \\
  & \hspace{7cm} 
  \underbrace{%
    \int \mbox{d} \bos{k}'\ \eem^{\iim (\bos{k}+\bos{k}')(\bos{r}_2-\bos{r}_1)}
  }_{%
    \left( 2 \pi\right)^3 \delta(\bos{r}_{2}-\bos{r}_1)
  } \ .
\end{align}
Using Eq.~(5.12b) of Ref.~\cite{sucherPhD1958}, we calculate the integral for the spherical angles, 
\begin{align}
  \int 
    \text{d} \hat{\bos{k}}\ 
    \phi_{\rm P}^\dagger(\bos{r}_1,\bos{r}_2) 
    \tilde{\bos{\alpha}}_1 \tilde{\bos{\alpha}}_2 
    \Lambda_{1-}(-\bos{k})\Lambda_{2-}(\bos{k}) 
    \phi_{\rm P}(\bos{r}_1,\bos{r}_2)
  = 
  4 \pi \frac{k^2}{6[{E^0}(k)]^2} 
  \langle %
    \bos{\sigma}_1 \bos{\sigma}_2 
  \rangle 
  \varphi^*(\bos{r}_1,\bos{r}_2)  \varphi(\bos{r}_1,\bos{r}_2) \ ,
\end{align}
where $\varphi$ is the non-relativistic wave function and  
$\langle \bos{\sigma}_1 \bos{\sigma}_2 \rangle$ has to be evaluated for the given spin state.
Substituting it back into the original integral,
\begin{align}
  \epsilon_{\mathrm{CB}}^{--}
  = 
  8 z^2_1 z^2_2 e^4 %
  \left\langle \varphi  \left| \delta(\bos{r}_{2}-\bos{r}_1) \right| \varphi \right\rangle 
  \left\langle  \bos{\sigma}_1 \bos{\sigma}_2  \right\rangle 
  \int_{0}^{\infty}  
    \text{d} k\  
    \frac{1}{6[E^0(k)]^2(2m + 2E^0(k))} \ ,
\end{align}
which is real, and thus, the $\text{Re}[...]$ can be omitted. 
\begin{align}
  \int_{0}^{\infty}  
    \text{d} k\  
    \frac{1}{6[E^0(k)]^2(2m + 2E^0(k))} 
  = \frac{\pi-2}{24  m^2 }
\end{align}
The integral can be evaluated analytically,
\begin{align}
  \epsilon_{\mathrm{CB}}^{--}
  = 
  \frac{z_1^2 z_2^2 e^4}{3m^2} 
  \left \langle\varphi  \left| \delta(\bos{r}_{2}-\bos{r}_1) \right| \varphi \right \rangle \left \langle  \bos{\sigma}_1 \bos{\sigma}_2 \right \rangle (\pi-2)  \ ,
\end{align}
which simplifies for singlet states, using $\left \langle  \bos{\sigma}_1 \bos{\sigma}_2 \right \rangle =-3$, to
\begin{align}
  \epsilon_{\mathrm{CB}}^{--}
  =
  -\frac{z_1^2 z_2^2 e^4 \,}{m^2} 
  \left \langle\varphi  \left| \delta(\bos{r}_{2}-\bos{r}_1) \right| \varphi \right \rangle  (\pi-2)  \ .
\end{align}
Then, with $\alpha=e^2$ and considering that $\langle \varphi | \delta(\bos{r}_2-\bos{r}_1)|\varphi\rangle$ carries an $m^3 \alpha^3$ factor, which combined with the $\alpha^2/m^2$ prefactor yields $m\alpha^5=\alpha^3 E_\mathrm{h}$,
\begin{align}
  \alpha^3 \epsilon_{3,\mathrm{CB}}^{--} 
  =
  -z_1^2 z_2^2 \alpha^3 \, 
  \left \langle\varphi  \left| \delta(\bos{r}_{2}-\bos{r}_1) \right| \varphi \right \rangle  (\pi-2) \ .
\label{Seq:epsBCfinalexpra}
\end{align}

\clearpage
\section{Convergence tables}
\begin{table}[h]
  \caption{%
With-pair (wp) and no-pair (np) Dirac-Coulomb and non-relativistic (nr) energies, in $\Eh$, as obtained from finite-basis computations reported in this work. Selected $\nb$ reference and $\naux$ auxiliary basis set data are shown. The energies correspond to a merged $\nb$ and $\naux$ basis set, so the total basis size is $N=\nb+\naux$.
    \label{tab:data}
  }
  \begin{tabular}{@{}ll r r r@{}}
    \hline\hline\\[-0.35cm] 
    \multicolumn{1}{l}{$\nb$} & 
    \multicolumn{1}{l}{$\naux$} &
    \multicolumn{1}{c}{$E_\text{nr}$} &
    \multicolumn{1}{c}{$E_\text{np}$} &
    \multicolumn{1}{c}{$E_\text{wp}$} \\
    \hline \\[-0.35cm]
\multicolumn{3}{l}{$\text{Ps}=\lbrace\text{e}^-,\text{e}^+\rbrace$:} \\
    30 & 10 & --0.249 999 999 996 78 & --0.249 997 552 766 9 & --0.249 997 554 234 8 \\
    30 & 20 & --0.249 999 999 999 53 & --0.249 997 552 778 1 & --0.249 997 554 256 3 \\
    40 & 20 & --0.249 999 999 999 88 & --0.249 997 552 780 1 & --0.249 997 554 259 7 \\
    50 & 30 & --0.249 999 999 999 92 & --0.249 997 552 780 1 & --0.249 997 554 259 6 \\
    $\infty^\dagger$ & &  --0.250 000 000 000 00& & \\
    \hline \\[-0.35cm]
\multicolumn{3}{l}{$\text{Mu}=\lbrace\text{e}^-,\mu^+\rbrace$:} \\
    30 & 10 & --0.497 593 472 890 81 & --0.497 600 026 229 05 & --0.497 600 026 230 17\\
    30 & 20 & --0.497 593 472 896 66 & --0.497 600 026 262 69 & --0.497 600 026 263 89\\
    40 & 20 & --0.497 593 472 911 84 & --0.497 600 026 270 86 & --0.497 600 026 272 04\\
    50 & 30 & --0.497 593 472 916 25 & --0.497 600 026 299 89 & --0.497 600 026 301 07\\
    $\infty^\dagger$ & &    --0.497 593 472 917 13 & & \\
    \hline \\[-0.35cm]
\multicolumn{3}{l}{$\text{H}=\lbrace\text{e}^-,\text{p}^+\rbrace$:}  \\
    30 & 10 & --0.499 727 839 669 37 & --0.499 734 619 507 95 & --0.499 734 619 507 95\\
    30 & 20 & --0.499 727 839 669 36 & --0.499 734 619 508 70 & --0.499 734 619 507 68 \\
    40 & 20 & --0.499 727 839 705 98 & --0.499 734 619 795 65 & --0.499 734 619 796 82\\
    50 & 20 & --0.499 727 839 709 47 & --0.499 734 619 841 35 & --0.499 734 619 841 38 \\
    50 & 30 & --0.499 727 839 710 95 & --0.499 734 619 841 92 & --0.499 734 619 841 94 \\
    $\infty^\dagger$ & &  --0.499 727 839 712 38 & & \\
    \hline \\[-0.35cm]
\multicolumn{3}{l}{$\mu\text{H}=\lbrace \mu^-,\text{p}^+\rbrace$:} &&  \\
    30 & 10 & --92.920 417 298 3 &  --92.920 891 299& --92.920 891 378 \\
    40 & 20 & --92.920 417 300 3 &  --92.920 891 313& --92.920 891 363 \\
    50 & 20 & --92.920 417 311 1 &  --92.920 891 315& --92.920 891 393\\
    60 & 30 & --92.920 417 311 1 &  --92.920 891 315& --92.920 891 396 \\
    $\infty^\dagger$ & &  --92.920 417 311 3 & & \\
    \hline\hline \\[-0.35cm]    
  \end{tabular}
    \begin{flushleft}
    $\dagger$: %
    Non-relativistic energy, $E_\text{nr}=- (1/2)(m_1m_2)/(m_1+m_2)$, using 
      $m_\mu= 206.768\ 283\ 0 \, m_\text{e}$, 
      $m_\text{p}= 1836.152\ 673\ 425\ 726  \, m_\text{e}$ \cite{MoNeTaTi25}.
  \end{flushleft}
\end{table}

\begin{table}[h]
  \caption{%
With-pair (wp) and no-pair (np) Dirac-Coulomb and non-relativistic (nr) energies, in $\Eh$, as obtained from finite-basis computations reported in this work. Selected $\nb$ reference and $\naux$ auxiliary basis set data are shown. The energies correspond to a merged $\nb$ and $\naux$ basis set, so the total basis size is $N=\nb+\naux$. 
    \label{tab:data}
  }
  \begin{tabular}{@{}rr r r r@{}}
    \hline\hline\\[-0.35cm] 
    \multicolumn{1}{l}{$\nb$} & 
    \multicolumn{1}{l}{$\naux$} &
    \multicolumn{1}{c}{$E_\text{nr}$} &
    \multicolumn{1}{c}{$E_\text{np}$} &
    \multicolumn{1}{c}{$E_\text{wp}$} \\
    \hline \\[-0.35cm]
\multicolumn{3}{l}{He (1S$_0$):} \\
    500  & 20  & --2.903 724 376 89 & --2.903 856 632 09 & --2.903 856 636 01  \\
    750  & 50  & --2.903 724 376 97 & --2.903 856 632 20 & --2.903 856 636 14 \\
    1000 & 50  & --2.903 724 377 00 & --2.903 856 632 24 & --2.903 856 636 19 \\
    1000 & 100 & --2.903 724 377 00 & --2.903 856 632 24 & --2.903 856 636 19 \\
    $\infty^\dagger$& & --2.903 724 377 03 & & \\
    \hline \\[-0.35cm]
    \multicolumn{3}{l}{He (2S$_0$):} \\
    200 & 20  & --2.145 974 010 956 & --2.146 084 755 696 & --2.146 084 755 990   \\
    300 & 50  & --2.145 974 043 846 & --2.146 084 789 253 & --2.146 084 789 545 \\
    400 & 50  & --2.145 974 045 693 & --2.146 084 791 129 & --2.146 084 791 423  \\
    400 & 100 & --2.145 974 045 700 & --2.146 084 791 152 & --2.146 084 791 461 \\
    $\infty^\dagger$& & --2.145 974 046 054 & & \\
    \hline \\[-0.35cm]
    \multicolumn{3}{l}{Li$^+$ (1S$_0$):} \\
    200 & 20  & --7.279 913 381 & --7.280 698 869 & --7.280 698 887  \\
    300 & 50  & --7.279 913 400 & --7.280 698 888 & --7.280 698 907  \\
    400 & 50 & --7.279 913 407  & -7.280 698 896  & --7.280 698 915 \\
    400 & 100 & --7.279 913 408  & --7.280 698 897  & --7.280 698 916 \\
    $\infty^\dagger$ & & --7.279 913 413 & & \\
    \hline \\[-0.35cm]
    \multicolumn{3}{l}{Be$^{2+}$ (1S$_0$):} \\
    100 & 20  & --13.655 565 911 & --13.658 257 255 & --13.658 257 300  \\
    200 & 50  & --13.655 566 228 & --13.658 257 596 & --13.658 257 650 \\
    300 & 50 & --13.655 566 234 & --13.658 257 602  & --13.658 257 657 \\
    300 & 100 & --13.655 566 234 & --13.658 257 604 & --13.658 257 660 \\
    $\infty^\dagger$ & & --13.655 566 238  & & \\
    \hline\hline \\[-0.35cm]    
  \end{tabular}
  \begin{flushleft}
  $\dagger$:
  Non-relativistic energy value (known to more digits) from Ref.~\citenum{Dr06}.
  \end{flushleft}
\end{table}


\end{document}